\documentclass[11pt]{article}

\usepackage[a4paper,margin=1in]{geometry}

\usepackage{amsmath,amssymb,amsthm,mathtools}
\usepackage{enumitem}
\usepackage{hyperref}
\usepackage{booktabs}
\usepackage{graphicx}
\usepackage{xcolor}
\usepackage[authoryear,round]{natbib}
\hypersetup{colorlinks=true,linkcolor=blue,citecolor=blue,urlcolor=blue}

\newcommand{\E}{\mathbb E}

\newcommand{\Var}{\operatorname{Var}}

\newcommand{\e}{\mathrm e}

\newcommand{\numplaceholder}[1]{%
\fbox{\parbox[c][2.0in][c]{0.92\textwidth}{%
\centering\textbf{Numerical exhibit placeholder}\\[0.7em]#1}}}

\theoremstyle{plain}
\newtheorem{theorem}{Theorem}[section]
\newtheorem{lemma}[theorem]{Lemma}
\newtheorem{proposition}[theorem]{Proposition}
\newtheorem{corollary}[theorem]{Corollary}

\theoremstyle{definition}

\theoremstyle{remark}
\newtheorem{remark}[theorem]{Remark}

\usepackage[affil-it]{authblk}

\title{Pooled Testing under Congestion:\\ Operational Scaling Laws for Pool-Size Selection}
\author[1]{Ioannis Dimitriou \footnote{ idimit@uoi.gr}\footnote{Corresponding author.}}
\author[2]{Johan van Leeuwaarden}
\affil[1]{Department of Mathematics, University of Ioannina, 45110, Ioannina, Greece}
\affil[2]{Department of Econometrics and Operations Research,
Tilburg University, Tilburg, The Netherlands}
\date{}

\begin{document}
\maketitle

\begin{abstract}
Classical pooled-testing rules choose pool sizes to reduce the number of tests. In practice, however, samples arrive over time, pools take time to form, testing capacity is finite, and positive pools create bursts of follow-up work. These operational features can make the test-efficient pool size far from the delay-efficient one.

We formulate pooled testing with stochastic arrivals and finite testing
capacity as a queueing system, and derive exact performance representations
and asymptotic scaling laws for fixed-batch Dorfman testing and a dynamic two-threshold policy. The analysis reveals three distinct design regimes. When test usage is the main concern, the classical Dorfman rule is recovered and the optimal pool size scales with the inverse square root of prevalence. When delay accumulated while the laboratory is busy testing another batch matters, positive pools create service-time variability that grows more rapidly with pool size than the mean retesting burden, and the optimal scale shifts to the inverse cube root of prevalence. When waiting to assemble a pool is itself costly, the optimal fixed pool size remains bounded even as prevalence becomes very small. Under severe capacity scarcity, congestion remains relevant at the classical Dorfman scale: the square-root law survives, but with a capacity-dependent constant.

Dynamic pooling separates two operational decisions that fixed pool sizes conflate. A lower threshold determines when routine testing should begin, while an upper threshold limits pooling when backlog is high. The lower threshold follows the cube-root scale, whereas the upper threshold primarily acts as congestion control. The results show how prevalence, capacity, and delay jointly determine pooling decisions and explain why operationally optimal pools can be much smaller than those suggested by test-efficiency calculations alone.\\


\end{abstract}

\section{Introduction}
\label{sec:introduction}

Pooled testing combines individual specimens and tests them jointly. If a pool
is negative, all specimens are cleared at once; if it is positive, additional
testing is used to identify the positive individuals. The idea dates to
Dorfman's 1943 proposal for syphilis screening \cite{dorfman1943} and has since
been used in blood-bank screening, HIV surveillance, hepatitis testing, malaria
monitoring, tuberculosis and influenza screening, and large-scale SARS-CoV-2
testing. These applications share the same basic statistical opportunity: when
prevalence is low, most pools are negative and a single assay can resolve many
individuals. Pooling can therefore economize on reagents, reduce test counts,
and expand effective diagnostic capacity
\cite{mutesa2021,cleary2021,barak2021}.

The classical design variable is the \emph{pool size}. Under Dorfman's two-stage
procedure, a pool of size $k$ is tested once; if it is positive, the $k$
individual specimens are retested. With individual positivity probability
$p_0$, expected tests per sample equal
\[
\frac{1}{k}+1-(1-p_0)^k.
\]
At low prevalence, $1-(1-p_0)^k\approx kp_0$, so minimizing testing effort
balances $1/k$ against $kp_0$. This produces Dorfman's familiar square-root
prescription $k_D\sim 1/\sqrt{p_0}$. The rule is simple and powerful: as
prevalence falls, larger pools become statistically attractive at a precisely
determined rate.

A diagnostic laboratory, however, does not operate on preassembled pools. It
receives a stream of samples. Samples must wait for enough companions to form a
pool; completed pools may wait for equipment; and positive pools create
additional testing that consumes capacity and makes service requirements
random. The pool-size decision therefore changes both the \emph{amount} of work
and the \emph{timing} of work. This distinction is consequential whenever the
value of a diagnostic result decays with delay. Blood products cannot be
released before screening is completed, infection-control decisions in
hospitals depend on timely surveillance, and the value of infectious-disease
testing for isolation, treatment, and contact management deteriorates as
turnaround times lengthen.

The tension is visible in practice. \cite{Cabrera2023} describe
pooled PCR testing in an acute-care hospital where demand exceeded the capacity
of individual testing. The reported data imply prevalence of approximately
$0.43\%$. At that prevalence, the square-root benchmark is about $15$ samples
and direct minimization of expected Dorfman tests gives a pool size of $16$.
The laboratory used pools of size $4$. From the viewpoint of test counts alone,
that choice looks conservative. From the viewpoint of a capacity-constrained
service operation, it need not be: the laboratory had to trade reagent savings
against formation delay, stochastic retesting workload, and turnaround time.

This paper develops an operational theory of that tradeoff. Our central
question is not whether pooling is beneficial, but \emph{how the optimal degree
of pooling scales when time and capacity matter}. The answer is not a single
correction to Dorfman's rule. Different operational objectives generate
different asymptotic balances and, as a consequence, different scaling laws.
Test-efficiency objectives retain the classical inverse-square-root scale.
Queueing-sensitive objectives produce an inverse-cube-root scale. Once the
time required to assemble a pool receives positive weight, the optimal
fixed-pool size remains bounded even as prevalence vanishes. A separate
scarce-capacity regime preserves the square-root exponent but changes its
leading constant.

Pooling is also inherently a \emph{dynamic batching decision}. Our basic policy
has two thresholds, $(l,k)$. The laboratory can start service once $l$ samples
are available. If more samples have accumulated by the time capacity becomes
available, it processes all waiting samples up to the cap $k$. Thus $l$ is a
release threshold that governs responsiveness in normal operation, whereas $k$
is a maximum pooling threshold that governs how aggressively backlog is
absorbed. The familiar fixed-pool policy is the special case $l=k$.

Two-threshold batching structures have appeared previously in related queueing
models. \cite{barlev2007bulk} optimize lower and upper batch
thresholds for an incomplete-identification bulk queue, and \cite{benjaafar2025congestion} report a two-threshold policy for a dynamic
extension of their pooled-testing model. Our use of $(l,k)$ builds on this
operational structure and provides an exact low-prevalence characterization of
the thresholds under complete identification.

The distinction between $l$ and $k$ is operationally important. A fixed pool
forces the laboratory to use the same number of samples when the system is
quiet and when it is congested. The $(l,k)$ policy separates those situations.
A laboratory can start relatively quickly under normal demand while still
using a larger pool when many samples are already waiting. The policy therefore
formalizes a practical distinction between \emph{responsiveness} and
\emph{batching aggressiveness}. At the same time, the fixed-batch model remains
analytically indispensable: it exposes the queueing mechanism most
transparently and admits an exact root representation from which the
heavy-traffic and scarce-capacity constants can be derived. We use the fixed
case to identify the scaling mechanisms and then return to $(l,k)$ to determine
which conclusions survive under dynamic batching.

The queueing analysis is exact. We embed the process at testing completions and
derive mean waiting time through renewal rewards and Little's law. For fixed
batches, the stationary generating function admits a root representation,
paralleling the exact $z$-transform approach in \cite{benjaafar2025congestion}. This representation allows us to identify
the heavy-traffic singularity directly: the root-sum term remains bounded,
whereas an explicit term diverges at rate $(1-\rho_k)^{-1}$, as $\rho_{k}\uparrow 1$, where $\rho_k$ is the utilization of the fixed-batch testing system. The resulting
heavy-traffic expression contains the familiar arrival- and service-variability
components, but it follows from the exact model rather than from an imposed
diffusion or Kingman approximation. For the state-dependent policy, a coupling
argument controls backlog by comparison with the fixed-batch system and yields
rigorous low-prevalence asymptotics.

The resulting scaling laws are summarized in
Table~\ref{tab:scaling-laws}. The table deliberately separates the
\emph{operational objective} from the asymptotic mechanism selecting the pool
size.
\begin{table}[t]
\centering
\caption{Operational low-prevalence scaling laws for pooled testing for 
delay-sensitive settings.%
}
\label{tab:scaling-laws}
\begin{tabular}{
    p{0.23\textwidth}
    p{0.37\textwidth}
    p{0.16\textwidth}
    p{0.14\textwidth}}
\toprule
\textbf{Objective}
&
\textbf{Dominant balance}
&
\textbf{Scaling}
&
\textbf{Result}
\\
\midrule

Testing effort
&
Pooled-test savings vs expected retesting
&
$k^\star\sim p_0^{-1/2}$
&
Theorem~\ref{thm:global-dorfman}
\\[0.7em]

Queueing delay
&
Batching gain vs variability generated by rare positive pools
&
$k^\star\sim p_0^{-1/3}$
&
Theorem~\ref{thm:global-queue-cuberoot}
\\[0.7em]

Testing plus queueing
&
Testing cost versus queueing consequences of positive-pool variability
&
$k^\star=\Theta(p_0^{-1/3})$
&
Theorem~\ref{thm:global-testing-queueing}
\\[0.7em]

Pool formation 
&
Benefits of pooling vs time required for batch assembly
&
$k^\star=\Theta(1)$
&
Theorem~\ref{thm:bounded-formation}
\\[0.7em]

Total turnaround 
&
Linear formation delay  prevents  optimal
pool size to diverge
&
$k^\star=\Theta(1)$
&
Corollary \ref{cor:bounded-turnaround}
\\
\bottomrule
\end{tabular}
\end{table}
The first row is the classical benchmark. Test efficiency balances the
pooled-test contribution $1/k$ against expected retesting of order $p_0k$,
yielding the square-root law. Queueing changes this balance because it is
sensitive not only to mean testing effort but also to variability. At low
prevalence a positive pool is rare, but when it occurs it creates a burst of
$k$ follow-up tests. The second moment of the testing cycle therefore contains
a term of order $p_0k^3$, which produces a queueing contribution of order
$p_0k^2$ after normalization by batch size. Balancing this term against the
batching gain $1/k$ leads to the inverse-cube-root scale. For clarity, the cube-root results refer to the busy-testing component
$W_{\mathrm{queue}}$ defined in Section \ref{sec:model}. This component is distinct from
the conventional post-formation queueing delay and from total turnaround
time, which also includes idle-time batch formation and in-testing delay.

The cube-root mechanism survives when testing effort and queueing are combined
in a weighted objective. A positive test cost changes the leading constant but
not the exponent. Formation delay changes the problem more fundamentally.
Under fixed batching, the expected delay experienced by tagged sample to assemble a pool is
$(k-1)/(2\lambda)$, where $\lambda$ is the sample arrival rate, and therefore increases linearly with $k$, independently of
how small prevalence becomes. Once formation delay has any fixed positive
weight, the optimal fixed-pool size can no longer diverge with decreasing
prevalence. The limiting recommendation is instead bounded.

Dynamic batching gives the cube-root result a more precise operational
interpretation. On the low-prevalence cube-root scale, the system starts
routine batches at the lower threshold with probability tending to one. The
release threshold therefore satisfies $l\sim p_0^{-1/3}$ within the
state-dependent scaling family, while the upper threshold $k$ remains a
separate congestion-control parameter. The operational prescription is thus
not simply to replace the classical fixed pool by a smaller fixed pool:
routine release and backlog absorption are distinct decisions.

Table~\ref{tab:scaling-laws} holds under fixed testing capacity. A different
phenomenon arises when capacity itself becomes scarce as prevalence decreases.
At the classical Dorfman scale $k=\Theta(p_0^{-1/2})$, fixed capacity makes
utilization vanish asymptotically, so ordinary congestion cannot generate a
first-order correction to Dorfman's recommendation. For congestion to remain
relevant on that scale, capacity must vary jointly with prevalence according to
$\mu/\lambda=\Theta(\sqrt{p_0})$, where $\mu$ is the service rate of one testing stage. Under the critical parameterization
$p_0=\varepsilon^2$ and $\mu/\lambda=c\varepsilon$, Theorem~\ref{thm:extendDorfman}
shows that the square-root exponent survives but its constant becomes
capacity dependent: 
\[
k^\star
\sim
\frac{x^\star(c)}{\sqrt{p_0}},
\qquad
x^\star(c)=c-\sqrt{c^2-3}.
\]
Thus congestion can affect pooling in two fundamentally different ways. Under
fixed capacity and a delay-sensitive objective, it contributes to a mechanism
that selects a new cube-root scale. Under scarce capacity, it remains
first-order within the classical square-root scale and changes the optimal
constant.

The paper builds on several strands of the pooled-testing literature. The
classical literature asks how pooling can reduce the number or cost of tests,
starting with Dorfman and extending to richer designs that account for
heterogeneous infection risks, imperfect tests, prevalence uncertainty,
dilution, and alternative assay technologies; see, among others,
\cite{wein1996dilution,aprahamian2019risk,aprahamian2020robust,
elhajj2022newborn,bish2024multidisease}
and the broader survey of \cite{aldridge2019information}.
Recent large-scale implementations confirm both the practical value of pooling
and the importance of prevalence and assay sensitivity in determining useful
pool sizes
\cite{abdalhamid2020pooling,bish2021robust,mutesa2021,cleary2021,barak2021}.
Our results retain Dorfman's two-stage protocol as the main benchmark, but the
object of optimization is different: the testing system is fed by stochastic
arrivals, so pool size also controls when service can start and how random
retesting work interacts with finite capacity.

A second strand places group testing explicitly in a queueing or dynamic
setting. \cite{barlev2007bulk} formulate a bulk-service queue in
which the batch size lies between lower and upper thresholds and optimize these
thresholds for an incomplete-identification group-testing problem. Their
$(m,M)$ structure is closely related to the release/cap logic of our $(l,k)$
policy, although contaminated groups are not individually resolved in their
model. \cite{bhimaraju2022scheduling} study the
scheduling of group tests over time. More generally, our embedded-chain
formulation belongs to the bulk-queue tradition; see, e.g.,~\cite{chaudhry1983bulk,bruneel1993discrete,janssen2005analytic}.

The paper closest to ours is \cite{benjaafar2025congestion}. They study a fixed-batch Dorfman system
with Poisson sample arrivals and stochastic testing times and make the key
operational point that maximizing throughput need not minimize the delay
experienced by samples. They derive an exact expression for expected delay,
characterize feasible pool sizes, and show numerically that the
throughput-maximizing pool can generate substantially larger delay than the
delay-minimizing pool. They also study a dynamic extension and report
numerically that its optimal policy has a two-threshold form. We view these
results as an important starting point for the present paper.

Our contribution is complementary to their finite-parameter optimization.
Rather than asking only which pool size minimizes delay for a given parameter
instance, we identify the asymptotic mechanisms that determine how the
operationally optimal pool size changes with prevalence and capacity. In
particular, we establish the cube-root queueing scale, characterize the
scarce-capacity regime in which a congestion-adjusted square-root law survives,
derive global optimizer scaling under alternative operational objectives, and
obtain exact asymptotics for a two-threshold $(l,k)$ policy. The last result
provides an analytical interpretation of the two-threshold structure observed
numerically by \cite{benjaafar2025congestion}: the lower threshold governs routine release,
while the upper threshold acts as a congestion cap. The two-threshold structure itself is therefore not claimed as a new policy class. Our contribution is its analytical characterization. In particular, we identify the low-prevalence scaling of the routine release threshold, explain the role of
the upper threshold as a congestion cap, and derive the associated asymptotic laws.

Other recent work emphasizes complementary operational considerations. \cite{lee2022} study pooling on high-throughput and point-of-care
platforms with attention to throughput and turnaround. \cite{zismanov2024} motivate combinatorial pooling partly by the delay
created by second-stage retesting. \cite{pagotto2025} examine
pooled PCR testing for hospital surveillance, where timely feedback is central.
\cite{augenblick2022pooled} emphasize repeated surveillance
and testing frequency, while \cite{long2026distributionally}
incorporate test errors and dependence into a distributionally robust pooling
model. Relative to this literature, our focus is on exact queueing foundations
and asymptotic design principles that reveal which operational mechanism
selects the pool-size scale.

The paper broadly makes four contributions. First, it provides an exact analytical characterization of the two-threshold batching policy, distinguishing the routine release threshold $l$ from the upper congestion-control threshold $k$. The policy structure is operationally equivalent to the $(M,Q)$ policy studied in \cite{benjaafar2025congestion}. The contribution here lies in deriving its stationary representation and asymptotic behavior. 
Second, it develops a unified embedded-chain and renewal-reward analysis for
state-dependent batching, with fixed batching recovered as a benchmark and an
exact root representation available in that case. This exact structure makes
it possible to identify the heavy-traffic and low-prevalence mechanisms without
relying on queueing approximations. Third, it establishes the operational scaling laws summarized in
Table~\ref{tab:scaling-laws}, including the square-root law for testing
effort, the cube-root law for the busy-testing component
$W_{\rm queue}$, the bounded regime under formation-sensitive and
total-turnaround objectives, and the capacity-dependent square-root
correction under scarce capacity. For the state-dependent policy, the
analysis further shows, within the considered cube-root scaling family,
that the cube-root scale governs the normal release threshold $l$,
whereas $k$ acts as a congestion-control parameter relevant in
atypical backlog states.
Fourth, the paper translates these analytical results into numerical examples
anchored by the prevalence reported in an acute-care hospital application. To
separate the operational mechanisms from an arbitrary physical time scale, the
capacity examples use the normalized ratio $\lambda/\mu=2$ and report delays
in units of the mean duration $1/\mu$ of one testing stage. The numerical
exhibits compare classical and operational policies, decompose turnaround time,
and show how the different scaling laws translate into finite-prevalence pool
sizes.

To illustrate the different operational scales, we calibrate a simple
example to the acute-care hospital implementation of \cite{Cabrera2023}. Their laboratory implemented four-in-one pooling
during a period in which testing demand exceeded available processing
capacity. In the first week after implementation, $10{,}904$ specimens
were processed using $2{,}914$ tests. With pools of size four, this
corresponds to $10{,}904/4=2{,}726$ first-stage pool tests. The remaining
$2{,}914-2{,}726=188$ tests correspond to $47$ positive pools followed
by four individual retests. Hence the observed positive-pool fraction is
$47/2726$.

Under the independent-positivity model, the probability that a pool of
four is positive equals $1-(1-p_0)^4$. We therefore calibrate the
individual positivity probability from
\(
1-(1-p_0)^4=\frac{47}{2726},
\)
which gives $p_0=0.0043385$. Cabrera et al.\ also report that pooling
increased laboratory capacity from approximately $3{,}000$ to more than
$6{,}000$ specimens per day. Motivated by this twofold capacity gap, we
use the normalized ratio $\lambda/\mu=2$ for the capacity illustration
below. This ratio should be viewed as an operational benchmark rather
than as a separate statistical estimate of $\lambda$ and $\mu$.

\begin{figure}[htp]
\centering
\includegraphics[width=\textwidth]{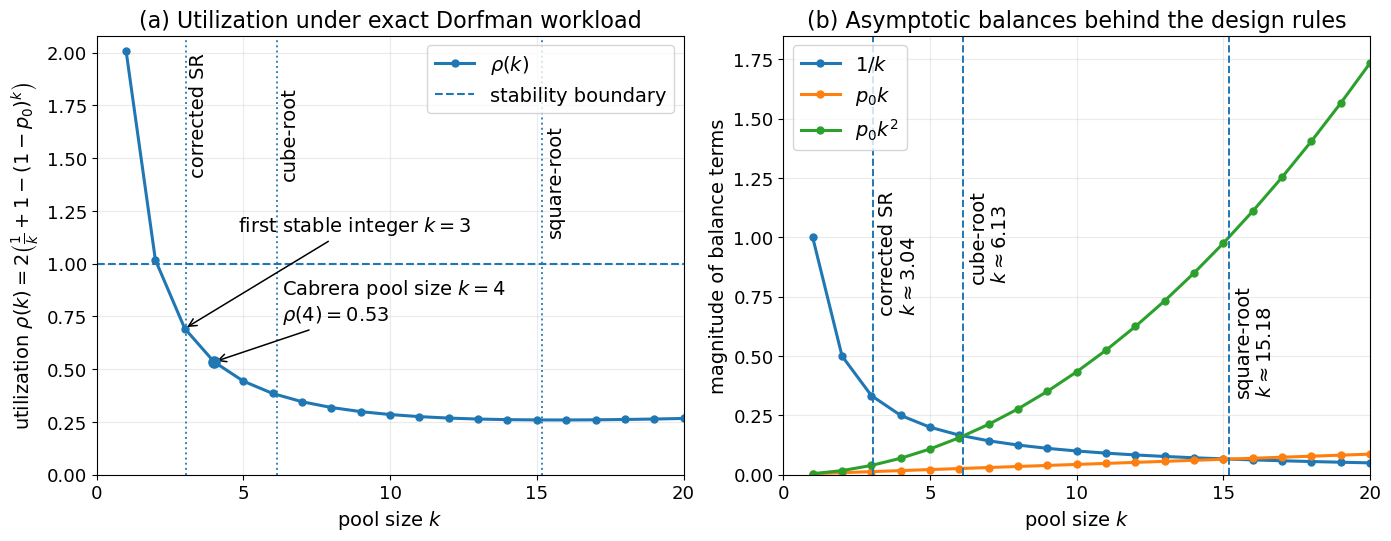}
\caption{Operational scaling laws under the \cite{Cabrera2023} calibration.
(a) Exact laboratory utilization as a function of pool size.
(b) Leading-order terms underlying the classical square-root,
scarce-capacity square-root, and queueing-induced cube-root design rules.}
\label{fig:intro_mechanism}
\end{figure}

Figure~\ref{fig:intro_mechanism} shows what the different design rules
mean at this calibration. The left panel plots the exact laboratory
utilization
\(
\rho_k:=\rho(k)
=
\frac{\lambda}{\mu}
\left(
\frac{1}{k}+1-(1-p_0)^k
\right)
\).
Without pooling, utilization is $\rho(1)=2.009$, so the laboratory is
unstable. Pooling rapidly reduces the workload: the first stable integer
pool size is $k=3$, while the pool size $k=4$ used by Cabrera et al.\
gives $\rho(4)=0.534$. Beyond this point, increasing the pool size
continues to reduce utilization initially, but the operational gains
quickly become much smaller.

The right panel explains why different objectives lead to different
recommended pool sizes. For small $p_0k$, the expected number of tests
per specimen is approximately
\(
{1}/{k}+p_0k
\).
The classical Dorfman rule therefore balances $1/k$ against $p_0k$,
giving $k\asymp p_0^{-1/2}$; here, $p_0^{-1/2}=15.18$. Under fixed
capacity with a queueing-sensitive objective, the variability generated
by positive pools introduces the larger-order term $p_0k^2$. Balancing
$1/k$ against $p_0k^2$ instead gives the cube-root scale
$k\asymp p_0^{-1/3}$, which equals $6.13$ in this example. Finally, in
the scarce-capacity regime of Theorem~\ref{thm:extendDorfman}, congestion
does not change the square-root exponent but does change its constant.
With $c=(\mu/\lambda)/\sqrt{p_0}=7.59$, the theorem prescribes a pool size of 
\(
(c-\sqrt{c^2-3})p_0^{-1/2}
=
3.04.
\)
Thus the same empirical calibration produces three quite different
operational scales: the classical efficiency rule favors pools around
$15$, delay-sensitive operation favors pools around $6$, while severe
capacity scarcity can push the preferred pool size down toward $3$.

The remainder of this article is structured as follows. Section \ref{sec:model} introduces the pooled-testing system and the dynamic $(l,k)$ batching policy. It also defines the sample-level delay decomposition used throughout the paper. Section \ref{sec:state-dependent-general} formalizes the model, the associated embedded queueing process, and develops the exact performance analysis. Using a Markov-renewal reward argument, we derive the pre-testing waiting time and its idle-time and busy-testing components. In Section \ref{sec:exact-asymptotics} we use these exact expressions to analyze operational delay in the heavy-traffic and low-prevalence regimes. These results identify the singular behavior near instability, the cube-root scaling of the renewal-reward queueing metric, and the corresponding extension to the dynamic $(l,k)$ policy. In Section \ref{sec:extendDorfman}, we consider a joint low-prevalence and scarce-capacity regime, and  investigate when congestion modifies the classical Dorfman square-root rule. Section \ref{sec:operational-objectives} turns the performance results into pool-size recommendations. We establish global optimizer results for testing effort, the renewal-reward queueing metric, combined testing and queueing costs, and formation-sensitive objectives. We further illustrate how they translate into pool-size choices at finite parameter values. Section \ref{sec:conclusion} concludes with a discussion of the operational consequences of the results. Additional numerical results and detailed proofs of the main results are provided in the Appendices.

\section{\texorpdfstring{Model and dynamic batching policy}{Model and dynamic batching policy}}
\label{sec:model}

Samples arrive according to a Poisson process with rate $\lambda$ and enter a
single capacity-constrained testing facility. A sample cannot be processed on
its own immediately upon arrival: it first has to be assigned to a testing
batch. For a batch of size $b$, let $S^{(b)}$ denote the duration of the
complete testing cycle, from the moment testing starts until every specimen in
the batch has received a definitive classification. The distribution of
$S^{(b)}$ is protocol specific. The queueing construction below only requires
these batch-level testing-time distributions; the explicit results later in
the paper specialize to Dorfman testing with exponential testing phases.

The key operational decision is therefore not simply a static pool size. It is
a \emph{batch-release rule}. Fix integers $1\le l\le k$. The lower threshold
$l$ is the smallest batch that the laboratory is willing to release, while the
upper threshold $k$ is the largest batch processed in a single testing cycle.
The decision is taken whenever the testing facility becomes available. If
fewer than $l$ samples are waiting, the laboratory deliberately waits until
$l$ samples have accumulated. If between $l$ and $k$ samples are waiting, it
starts immediately and processes all of them. If at least $k$ samples are
waiting, it starts immediately with a full batch of size $k$ and leaves the
remaining samples in queue.

This distinction is operationally important. Waiting for additional samples
when the system is quiet creates \emph{formation delay}; using additional
samples that are already waiting does not. The lower threshold $l$ therefore
controls responsiveness in normal operation. The upper threshold $k$ controls
how aggressively the laboratory can absorb backlog once congestion has already
created a queue. A fixed-pool model forces these two decisions to coincide.
Our policy separates them. Fixed batching is recovered as the special case
$l=k$.

The policy is deliberately simple. It does not require forecasting future
arrivals or solving a dynamic program at each decision epoch. It uses only the
queue length observed when testing capacity becomes available and translates
that state into a batch size. This makes $(l,k)$ both analytically tractable
and operationally interpretable: $l$ is a release threshold and $k$ is a
congestion cap.

For a fixed batch size $k$, the natural traffic intensity is
$\rho_k=\lambda\E[S^{(k)}]/k$. Stability requires $\rho_k<1$, because a
testing cycle processes $k$ samples while, on average, $\lambda\E[S^{(k)}]$
new samples arrive during that cycle. For the state-dependent $(l,k)$ policy,
the exact embedded-chain drift is governed by the full-batch state: whenever
the queue is large, the system repeatedly serves batches of size $k$. The
corresponding stability condition is therefore again
$\lambda\E[S^{(k)}]<k$; see Section~\ref{sec:state-dependent-general}. We use
the notation $\rho_k$ both for the fixed-batch benchmark and for the dynamic
system near this full-batch capacity boundary.

For a tagged sample, let $\bar W$ denote the time from its arrival until the batch containing it enters testing. We decompose this pre-testing delay according to the state of the testing facility as $\bar W=W_{\rm idle}+W_{\rm queue}$. In particular, $W_{\rm idle}$ measures the portion of the pre-testing waiting time accumulated while the testing facility is idle (i.e., idle-time batch formation), whereas $W_{\rm queue}$ is the portion accumulated while the facility is busy testing another batch. It includes both the case in which the tagged batch has already been formed but is waiting behind an earlier batch, and the case in which the tagged batch is still being formed while the laboratory is busy. Therefore, the turnaround time is given by $W=W_{\rm idle}+W_{\rm queue}+W_{\rm test}$, where $W_{\rm test}$ is the time from the beginning of the tagged batch's testing cycle until the final classification of the sample. The distinction between the first two components matters. A large pool can save tests while increasing the time needed to form a batch; conversely, when backlog is already present, a large batch can be formed without additional formation delay but may create a long or highly variable testing cycle.
\begin{remark}[Relation to the decomposition in \cite{benjaafar2025congestion}]
The pre-testing waiting time $\bar W$ considered here coincides with the total pre-testing waiting time of a tagged sample analyzed in \cite{benjaafar2025congestion}. However, the decompositions are different. In their formulation, $\bar W=W_{\rm form}+W_q$, where $W_{\rm form}$ is the time required for the tagged sample's batch to be formed and $W_q$ is the subsequent waiting time of the formed batch before its testing begins. In the fixed-batch model, $\E[W_{\rm form}]=\frac{k-1}{2\lambda}$, $\E[W_q]=\E[\bar W]-\frac{k-1}{2\lambda}$. Contrary to that case, the decomposition used in this paper partitions the pre-testing waiting time according to the state of the testing facility. Thus, although the two formulations yield the same total pre-testing waiting time $\bar W$, their queueing components are defined differently and do not coincide. In particular, the $W_{\rm queue}$ used in this paper includes every part of the pre-testing waiting time during which the laboratory is busy testing another batch, including periods during which the tagged batch is still being formed. The quantity $W_q$ in \cite{benjaafar2025congestion} refers only to the waiting of the tagged batch after its formation. 
\end{remark}

Under fixed batching, all three consequences are tied to the same parameter
$k$. Under $(l,k)$, routine formation is governed primarily by $l$, whereas
$k$ matters when backlog is present. One of the central results later in the
paper is that, in the low-prevalence delay-sensitive regime, the system chooses
a batch of size $l$ with probability tending to one. Thus the asymptotic
pool-sizing rule becomes a rule for \emph{routine release}, while $k$ remains
available as a separate congestion-control parameter.

The analysis begins with the general $(l,k)$ system because this is the
primitive operational policy. We first derive exact steady-state expressions
without imposing a particular testing protocol or asymptotic regime. We then
specialize to fixed-batch Dorfman testing, where the kernel admits a convenient
root representation. That benchmark exposes the heavy-traffic and
low-prevalence mechanisms particularly clearly. Finally, we return to the
state-dependent policy and show how those mechanisms translate into a release
threshold and a congestion cap.

\section{\texorpdfstring{Dynamic batching and exact performance}{Dynamic batching and exact performance}}
\label{sec:state-dependent-general}

We now formalize the $(l,k)$ policy introduced in Section~\ref{sec:model} and
derive its exact steady-state performance. The analysis is written for general
testing-cycle distributions $S^{(b)}$ and therefore separates the queueing
mechanism from the details of a particular screening protocol.

The natural observation epochs are testing completions. At such an epoch the
previous batch has been completely resolved and the laboratory is again
available, so the only state variable needed for the next batching decision is
the number of samples waiting. This turns the continuous-time system into a
one-dimensional embedded chain. Let $T_n$ denote the epoch at which the $n$th testing cycle is
completed, and define $Q_n$ be the queue length immediately
after that completion. If $Q_n=i$, the next batch size is

\[
f(i)
=
\begin{cases}
l, & 0\le i<l,\\
i, & l\le i<k,\\
k, & i\ge k.
\end{cases}
\]

Hence $l$ is the release threshold and $k$ is the capacity cap. Below $l$ the
system waits for additional samples; between $l$ and $k-1$ it clears the entire
queue; and at or above $k$ it serves a full batch. This distinction later lets
us identify which threshold governs normal operation and which threshold is
used primarily to absorb congestion.

For each possible batch size \(b\in\{l,l+1,\ldots,k\}\), let
\(S^{(b)}\) denote the duration of the testing procedure for a batch of
size \(b\). We assume that the sequence of testing times is independent of
the arrival process and that the distribution of \(S^{(b)}\) depends on the
batch size \(b\), but not on the history of the system.

Let \(A^{(b)}\) denote the number of arrivals during a testing cycle with
batch size \(b\). Conditional on \(S^{(b)}=s\), we have $A^{(b)}\mid S^{(b)}=s
\sim
\operatorname{Poisson}(\lambda s)$. Therefore, the pgf of \(A^{(b)}\) is 
\[
A^{(b)}(z)
=
\E[z^{A^{(b)}}]
=
\E\left[
\exp\{-\lambda(1-z)S^{(b)}\}
\right].
\]

The embedded waiting-queue process satisfies $Q_{n+1}
=
\bigl(Q_n-f(Q_n)\bigr)^+
+
A_{n+1}^{(f(Q_n))}.$
Let
\[
\Pi(z)
=
\sum_{i=0}^{\infty}\pi_i z^i,
\]
be the stationary pgf of \(Q_n\). Conditioning on \(Q_n\), and rearranging yields 
\begin{equation}
\Pi(z)
=
\frac{
z^k
\left[
A^{(l)}(z)\sum_{i=0}^{l-1}\pi_i
+
\sum_{i=l}^{k-1}\pi_i A^{(i)}(z)
\right]
-
A^{(k)}(z)\sum_{i=0}^{k-1}\pi_i z^i
}{
z^k-A^{(k)}(z)
}.
\label{eq:general-state-dependent-pgf}
\end{equation}

The central object is the kernel $K_k(z)=z^k-A^{(k)}(z)$. Its form reveals
an important structural property of the policy. State dependence enters the
numerator of \eqref{eq:general-state-dependent-pgf} through the release
decisions taken in the boundary states $0,\ldots,k-1$. The denominator,
however, is determined entirely by the dynamics when the queue is large enough
to trigger a full batch of size $k$. Hence the stability and root structure are
governed by the congestion cap, while the lower threshold changes the boundary
conditions. This separation is what makes the root analysis reusable across
batching rules and later allows us to distinguish normal release from
congestion control. Under the stability
condition
$\lambda \E[S^{(k)}]< k$,
the embedded chain is positive recurrent.

Let \(\xi_1,\ldots,\xi_{k-1}\) denote the roots of
$z^k-A^{(k)}(z)=0$
strictly inside the unit disk. Analyticity of \(\Pi(z)\) implies that the
numerator in \eqref{eq:general-state-dependent-pgf} vanishes at these
roots. Hence the boundary probabilities satisfy
\[
\sum_{i=0}^{l-1}
\pi_i
\left(
A^{(l)}(\xi_m)-\xi_m^i
\right)
+
\sum_{i=l}^{k-1}
\pi_i
\left(
A^{(i)}(\xi_m)-\xi_m^i
\right)
=
0,
\qquad
m=1,\ldots,k-1.
\]
The final equation is normalization, equivalently the stationary flow
balance
\begin{align}
&
\sum_{i=0}^{l-1}\pi_i l
+
\sum_{i=l}^{k-1}\pi_i i
+
k\left(1-\sum_{i=0}^{k-1}\pi_i\right)
\nonumber\\
&\qquad=
\lambda
\left[
\sum_{i=0}^{l-1}\pi_i\left(\frac{l-i}{\lambda} +\E[S^{(l)}]\right)
+
\sum_{i=l}^{k-1}\pi_i \E[S^{(i)}]
+
\left(1-\sum_{i=0}^{k-1}\pi_i\right)\E[S^{(k)}]
\right].
\label{eq:general-boundary-normalization}
\end{align}

The stationary queue-length distribution is not itself the performance
measure of interest. The operational quantity is the delay experienced by a
sample. We therefore convert the embedded-chain description into sample-level
waiting time by a renewal-reward argument. This avoids having to construct a
larger continuous-time Markov chain that keeps track simultaneously of queue
length, batch formation, and the phase of the testing procedure. 

For the $n$th cycle $[T_n,T_{n+1})$ define
\begin{displaymath}
    C_n:=T_{n+1}-T_n,\qquad D_n:=\int_{T_n}^{T_{n+1}}X(t)\,\mathrm{d}t,
\end{displaymath} 
where $X(t)$ is the number of samples waiting to enter testing at time $t$. A cycle contains a formation period only when the post-completion queue is below $l$;
otherwise testing starts immediately. Conditional on $Q_n=i$, its mean is

\[
\E[C_n\mid Q_n=i]
=
\begin{cases}
\dfrac{l-i}{\lambda}+\E[S^{(l)}],
& 0\le i<l,\\[3mm]
\E[S^{(i)}],
& l\le i<k,\\[2mm]
\E[S^{(k)}],
& i\ge k.
\end{cases}
\]
Therefore, if $C$ denotes a generic cycle length under the stationary
law,
\begin{align}
\E[C]
=&
\sum_{i=0}^{l-1}
\pi_i
\left(
\frac{l-i}{\lambda}
+
\E[S^{(l)}]
\right)
+
\sum_{i=l}^{k-1}
\pi_i\E[S^{(i)}]
+
\left(
1-\sum_{i=0}^{k-1}\pi_i
\right)\E[S^{(k)}].
\label{eq:general-cycle-mean}
\end{align}

Under the state-dependent $(l,k)$ policy, the cycle variables are not generally iid, because their distributions depend on $Q_n$. However, conditional on $Q_n$, the next cycle is generated independently of the previous history. Thus, the sequence $(Q_n,C_n,D_n)$ has a Markov-renewal structure; see \cite{Cinlar1969}. Since $Q_n$ is positive recurrent, the corresponding ergodic reward theorem gives
\begin{displaymath}
    \lim_{t\to\infty}\frac{1}{t}\int_0^t X(s)\,\mathrm{d}s=\frac{\E[D]}{\E[C]},
\end{displaymath}
where $D$ denotes the corresponding stationary cycle reward. Then, Little's law yields
\begin{equation}
\E[\bar W]
=
\frac{\E[D]}{\lambda\E[C]}
=
\frac{
\E\left[
\int_{T_n}^{T_{n+1}}X(t)\,\mathrm{d}t
\right]
}{
\lambda\E[C]
}.
\label{eq:general-wait-renewal}
\end{equation}

The expected accumulated waiting time in one cycle is
\begin{align}
\E\left[
\int_{T_n}^{T_{n+1}}X(t)\,\mathrm{d}t
\right]
=&
\sum_{i=0}^{l-1}
\pi_i
\left[
\sum_{m=0}^{l-i-1}
\frac{i+m}{\lambda}
+
\frac{\lambda}{2}\E[(S^{(l)})^2]
\right]
+
\sum_{i=l}^{k-1}
\pi_i
\frac{\lambda}{2}\E[(S^{(i)})^2]
\nonumber\\
&+
\sum_{i=k}^{\infty}
\pi_i
\left[
(i-k)\E[S^{(k)}]
+
\frac{\lambda}{2}\E[(S^{(k)})^2]
\right].
\label{eq:general-wait-reward}
\end{align}

The first term in the right-hand side of \eqref{eq:general-wait-reward} corresponds to the time accumulated while the facility is idle and the tagged batch is being formed. We therefore define
\begin{displaymath}
    \E[W_{\rm idle}(l,k)]:=\frac{1}{\lambda\E[C]}\sum_{i=0}^{l-1}\pi_i\sum_{m=0}^{l-i-1}\frac{i+m}{\lambda}.
\end{displaymath}
The remaining terms correspond to waiting accumulated while the facility is busy testing another batch. We define
\begin{align}
    E[W_{\rm queue}(l,k)]:=\frac{1}{\lambda\E[C]}\Bigg[\frac{\lambda}{2}\sum_{i=0}^{l-1}\pi_i
\E[(S^{(l)})^2]+\frac{\lambda}{2}\sum_{i=l}^{k-1}\pi_i\E[(S^{(i)})^2]\nonumber\\\qquad\qquad\qquad+\sum_{i=k}^{\infty}\pi_i\left((i-k)\E[S^{(k)}]+\frac{\lambda}{2}\E[(S^{(k)})^2]\right)\Bigg].\label{eq:general-queue-reward}
\end{align}
Consequently, $\E[\bar W(l,k)]=\E[W_{\rm idle}(l,k)]+\E[W_{\rm queue}(l,k)]$. The first quantity is the portion of the tagged sample's pre-testing delay accumulated while the laboratory is idle and its batch is being completed. The second quantity is the portion accumulated while the laboratory is busy testing another batch. Consequently, it includes both the waiting of a batch that has already been formed and the part of the tagged batch's formation period during which another batch is being tested.

Thus, \eqref{eq:general-wait-renewal} gives the mean waiting time before
entering testing. The reward decomposition also gives a useful interpretation.
States below $l$ generate formation delay, states above $k$ generate an
explicit backlog term, and all testing cycles generate a second-moment term
through arrivals that occur while capacity is occupied. The last component is
the source of the rare-event variability effect that drives the cube-root
results later in the paper.

To include the time spent in the testing procedure itself, let $\tau_b
=
\E[T_{\mathrm{test}}\mid \text{batch size }b]$ 
be the mean time from the start of testing until an arbitrary sample in a
batch of size \(b\) receives its final result. Let $B_n=f(Q_n)$ denote the number of samples released in the $n$th testing cycle. In stationarity, let $B$ be a generic cycle batch size. Hence, the expected number of samples processed in one cycle is
\[
\E[B]
=
\sum_{i=0}^{l-1}\pi_i l
+
\sum_{i=l}^{k-1}\pi_i i
+
k\left(1-\sum_{i=0}^{k-1}\pi_i\right),
\]
and the expected accumulated in-testing delay over one cycle is
\begin{align*}
\E[R_{\mathrm{test}}]
=&
\sum_{i=0}^{l-1}\pi_i l\tau_l
+
\sum_{i=l}^{k-1}\pi_i i\tau_i
+
k\left(1-\sum_{i=0}^{k-1}\pi_i\right)\tau_k.
\label{eq:general-test-reward}
\end{align*}
Since \(\E[B]=\lambda\E[C]\) in stationarity,
\[
\E[T_{\mathrm{test}}]
=
\frac{\E[R_{\mathrm{test}}]}{\lambda\E[C]}.
\]
Therefore the mean total delay is
\[
\E[W]
=
\frac{
\E\left[
\int_{T_n}^{T_{n+1}}X(t)\,\mathrm{d}t
\right]
+
\E[R_{\mathrm{test}}]
}{
\lambda\E[C]
}.
\]

This expression is exact and already contains the central operational tradeoff
of the paper. The release threshold enters through the time needed to complete
a batch when the system is below $l$. The congestion cap enters through the
backlog states in which a full batch of size $k$ is removed. The testing
protocol enters through the first two moments of $S^{(b)}$ and through the
sample-level resolution time $\tau_b$. In particular, the queue reacts to the
\emph{complete testing-cycle distribution}, not merely to its mean. This is why
a protocol with the same expected number of tests can have a very different
delay profile if it produces more variable follow-up workload.

The reward representation also clarifies why the subsequent asymptotics can be
modular. To study a different group-testing protocol one does not need to
rebuild the queue from scratch. One replaces the family of testing-cycle
distributions $S^{(b)}$ and the corresponding $\tau_b$, while the embedded
batching mechanism and renewal-reward structure remain the same. Dorfman
testing is the main application because its rare positive-pool workload makes
the interaction between prevalence, variability, and congestion especially
transparent.

\subsection{The fixed-batch Dorfman model as a special case}
\label{subsec:fixed-batch-dorfman-from-general}

The fixed-batch Dorfman system used in the asymptotic analysis follows directly
from the preceding construction by setting $l=k$. Then every testing cycle uses
a batch of size $k$, possibly after waiting for the batch to be completed.
Although fixed batching is less flexible operationally, it is the right
analytical benchmark: the policy has a single design parameter and the
resulting kernel can be studied exactly, allowing the source of the different
scaling laws to be isolated transparently. This benchmark is closely related
to the fixed-batch queueing model of \cite{benjaafar2025congestion}, who derive expected delay by a
$z$-transform/root approach and optimize the resulting finite-parameter
problem. We retain that operational benchmark but use it to obtain exact
heavy-traffic and low-prevalence asymptotics before returning to the dynamic
policy. Hence
$f(i)=k$, $i\ge0$,
and
$S^{(f(Q_n))} = S^{(k)}$. For Dorfman testing, the testing-cycle duration satisfies
\[
S^{(k)}
\sim
\begin{cases}
\operatorname{Exp}(\mu),
& \text{with probability }1-p_k,\\[1mm]
\operatorname{Erlang}(k+1,\mu),
& \text{with probability }p_k,
\end{cases}
\]
where
$p_k = 1-(1-p_0)^k$.
Therefore,
\begin{displaymath}
        \E[S^{(k)}]
=
\frac{1+k p_k}{\mu}, \qquad
      \E[(S^{(k)})^2]
=
\frac{2(1-p_k)}{\mu^2}
+
\frac{(k+1)(k+2)p_k}{\mu^2}.  
\end{displaymath}

The general state-dependent recursion reduces to
$Q_{n+1} = (Q_n-k)^+ + A_{n+1}$,
and \eqref{eq:general-state-dependent-pgf} reduces to
\[
\Pi(z)
=
\frac{
A(z)\sum_{i=0}^{k-1}\pi_i(z^k-z^i)
}{
z^k-A(z)
},
\]
where $A(z)$ is the pgf of the number of arrivals during one testing cycle. We now derive the mean delay from the general expression. Since \(l=k\),
the cycle mean in \eqref{eq:general-cycle-mean} becomes
\[
\E[C]
=
\frac{1}{\lambda}
\sum_{i=0}^{k-1}\pi_i(k-i)
+
\E[S^{(k)}].
\]
In stationarity, exactly \(k\) samples enter testing in each cycle, and
therefore
$\lambda\E[C]=k$.
Thus
$\E[C] = \frac{k}{\lambda}$.

Similarly, the accumulated waiting time before entering testing becomes
\begin{align}
\E\left[
\int_{T_n}^{T_{n+1}}X(t)\,\mathrm{d}t
\right]
=&
\frac{1}{2\lambda}
\sum_{i=0}^{k-1}
\pi_i(k-i)(k+i-1)
+
\E[S^{(k)}]
\sum_{i=k}^{\infty}\pi_i(i-k)
+
\frac{\lambda}{2}\E[(S^{(k)})^2].
\label{eq:fixed-dorfman-waiting-reward}
\end{align}
Dividing by \(\lambda\E[C]=k\), we obtain
\begin{equation}
\E[\bar W]
=
\frac{1}{k}
\left\{
\frac{1}{2\lambda}
\sum_{i=0}^{k-1}
\pi_i(k-i)(k+i-1)
+
\E[S^{(k)}]
\sum_{i=k}^{\infty}\pi_i(i-k)
+
\frac{\lambda}{2}\E[(S^{(k)})^2]
\right\}.
\label{eq:fixed-dorfman-wait-boundary}
\end{equation}

For fixed batch size $k$, equation \eqref{eq:fixed-dorfman-wait-boundary} can be decomposed as $\E[\bar W]=\ E[W_{\rm idle}]+\E[W_{\rm queue}]$,
where
\begin{equation}
\begin{array}{rl}
     \E[W_{\rm idle}]=& \frac{1}{2\lambda k}\sum_{i=0}^{k-1}\pi_i(k-i)(k+i-1),\vspace{2mm}\\
     \E[W_{\rm queue}]=&\frac{1}{k}\left[\E[S^{(k)}]\sum_{i=k}^{\infty}\pi_i(i-k)+\frac{\lambda}{2}\E[(S^{(k)})^2]
\right]. 
\end{array}\label{eq:delay-comp}
\end{equation}
The testing delay of an arbitrary sample in a Dorfman batch of size \(k\)
is $\tau_k
=
\frac{1}{\mu}
+
p_k\frac{k+1}{2\mu}.$ Therefore the mean total delay is
$\E[W] = \E[\bar W] + \tau_k$.

For asymptotic analysis it is useful to eliminate the boundary probabilities
and express the waiting time through the kernel roots. Let $\xi_1,\ldots,\xi_{k-1}$
be the roots strictly inside the unit disk of
$z^k-A(z)=0$.
For the fixed-batch model, the standard root representation gives
\begin{equation}
\E[\bar W]
=
\frac{1}{\lambda}
\sum_{m=1}^{k-1}
\frac{1}{1-\xi_m}
+
\frac{
\lambda \E[(S^{(k)})^2]
-
(k-1)\E[S^{(k)}]
}{
2(k-\lambda\E[S^{(k)}])
}.
\label{eq:fixed-dorfman-inside-root-wait}
\end{equation}
Consequently,
\begin{equation}
\E[W]
=
\frac{1}{\lambda}
\sum_{m=1}^{k-1}
\frac{1}{1-\xi_m}
+
\frac{
\lambda \E[(S^{(k)})^2]
-
(k-1)\E[S^{(k)}]
}{
2(k-\lambda\E[S^{(k)}])
}
+
\frac{1}{\mu}
+
p_k\frac{k+1}{2\mu}.
\label{eq:fixed-dorfman-inside-root-total}
\end{equation}


The exact expressions above will be used in two complementary ways. The
boundary-probability form \eqref{eq:fixed-dorfman-wait-boundary} is convenient
for low-prevalence analysis because it separates the second-moment and backlog
contributions. The root form \eqref{eq:fixed-dorfman-inside-root-wait} is
particularly useful in heavy traffic because it isolates the singular
denominator $k-\lambda\E[S^{(k)}]$. Keeping both representations, i.e., the renewal-reward representation \eqref{eq:fixed-dorfman-wait-boundary} and the equivalent inside-root representation \eqref{eq:fixed-dorfman-inside-root-wait}, is therefore not redundant: they expose different operational mechanisms of the same exact
queue. For completeness, an alternative outside-root representation of the fixed-batch formulas is given in Appendix \ref{app:outside-roots}. This representation is not used in the subsequent asymptotic analysis.

\section{Exact asymptotics for operational delay}
\label{sec:exact-asymptotics}

The exact queueing formulas now become the basis for the asymptotic analysis.
The purpose of this section is to identify the mechanisms that later determine
optimal pool sizes. Two limits are especially revealing. Heavy traffic shows
how testing-time variability is amplified when capacity is nearly exhausted.
Low prevalence shows how that variability itself changes with the pool size
even when utilization is small.

We first identify the heavy-traffic singularity directly from the root
representation and then derive the low-prevalence queueing scale for fixed and
state-dependent batches. This order matters. The square-root and cube-root
regimes developed later are not heuristic substitutions into a generic
queueing approximation; they are consequences of exact first-order limits. A
Kingman-type expression appears only after the exact heavy-traffic result, as
an operational interpretation of its variability terms.

\subsection{Heavy traffic from the exact root representation}

For the fixed-batch model, recall the exact waiting-time representation
\eqref{eq:fixed-dorfman-inside-root-wait}.  Write $\rho_k=\frac{\lambda\E[S^{(k)}]}{k}$, $c_S^2(k)=\frac{\Var(S^{(k)})}{\E[S^{(k)}]^2}$.
The second explicit term in \eqref{eq:fixed-dorfman-inside-root-wait} can be
rewritten as
\begin{equation}
\frac{\lambda \E[(S^{(k)})^2]-(k-1)\E[S^{(k)}]}
     {2(k-\lambda\E[S^{(k)}])}
=
\frac{\E[S^{(k)}]}{2}
\left[
\frac{\rho_k c_S^2(k)+1/k}{1-\rho_k}-1
\right].
\label{eq:exact-ht-explicit-term}
\end{equation}
The next proposition shows that this term contains the complete first-order
heavy-traffic singularity. Its proof is given in Appendix~\ref{app:proof-exact-heavy-traffic}.

\begin{proposition}[Exact fixed-batch heavy-traffic limit]
\label{prop:exact-heavy-traffic}
Fix $k$, $p_0$, $\mu$, and let $\lambda\uparrow k/\E[S^{(k)}]$, so that $\rho_k\uparrow1$,
\begin{equation}
\E[W_{\rm queue}(k)]
\sim
\frac{\E[S^{(k)}]}{2}
\frac{c_S^2(k)+1/k}{1-\rho_k}.
\label{eq:exact-heavy-traffic}
\end{equation}
\end{proposition}

\paragraph{Proof sketch.}
The exact root representation separates the waiting time into three pieces. The idle contribution and the root-sum contribution remain bounded as $\rho_k\uparrow1$, whereas the explicit second-moment term diverges at rate $(1-\rho_k)^{-1}$. Isolating that single singular term yields \eqref{eq:exact-heavy-traffic}. The complete argument, including the uniform bound on the roots, is given in Appendix~\ref{app:proof-exact-heavy-traffic}.

The exact limit also identifies the source of the two variability terms. Let
$T_A^{(k)}$ denote the time required for $k$ Poisson arrivals. Then
$T_A^{(k)}\sim\mathrm{Erlang}(k,\lambda)$ and $c_A^2(k)
:=
\frac{\Var(T_A^{(k)})}{\E[T_A^{(k)}]^2}
=
\frac1k.$
This notation is deliberately distinct from $A^{(k)}$, which elsewhere denotes
the number of arrivals during a testing cycle and is mixed Poisson. Thus the
exact heavy-traffic coefficient has the familiar GI/G/1 combination
$c_A^2+c_S^2$. This correspondence explains the form of the limit, but it is
not an approximation step: the singular term in
\eqref{eq:exact-heavy-traffic} has already been established from the exact
root representation.

Combining batching, queueing, and in-testing delay gives the transparent
three-delay decomposition
\begin{equation}
\E[W(k)]
\approx
\E[W_{\rm idle}(k)]
+
\frac{\rho_k}{1-\rho_k}
\frac{c_A^2(k)+c_S^2(k)}{2}\E[S^{(k)}]
+
\tau_k,
\label{eq:three-delay-interpretation}
\end{equation}
where $c_A^2(k)=1/k$.  We use \eqref{eq:three-delay-interpretation} only for
operational intuition.  The scaling results below are established from the
exact reward and root formulas.

For Dorfman testing, $\tau_k=\frac1\mu+p_k\frac{k+1}{2\mu}$, $\E[S^{(k)}]=\frac{1+kp_k}{\mu}$
and therefore $\rho_k
=
\frac{\lambda}{\mu}\left(p_k+\frac1k\right).$
The quantity $p_k+1/k$ is exactly the expected testing workload generated per
arriving sample: $1/k$ pooled tests per sample plus the expected individual
retesting contribution $p_k$.

\subsection{Low-prevalence queueing asymptotics: fixed batches}
\label{subsec:exact-cuberoot-fixed}

We next turn to the regime in which prevalence tends to zero while testing
capacity is fixed.  The following lemma identifies the exact order of the
queueing component without invoking a diffusion approximation.

\begin{lemma}[Fixed-batch queueing order]
\label{lem:fixedBSQ}
Assume $p_0\downarrow0$, $k\to\infty$, and $k^2p_0\to0$.  For the
fixed-batch Dorfman system,
\begin{equation}
\E[W_{\rm queue}(k)]
=
\Theta\!\left(\frac1k+k^2p_0\right).
\label{eq:fixed-queue-order}
\end{equation}
\end{lemma}

\paragraph{Proof sketch.}
Starting from the exact renewal-reward expression, the second-moment term is of order $1/k+k^2p_0$. The remaining excess-backlog term can be controlled through the Lindley recursion for $(Q_n-k)^+$ and Kingman's inequality; under $k^2p_0\to0$ it is of strictly smaller order. Thus the service-time second moment determines the queueing scale. Appendix~\ref{app:proof-fixed-batch-order} contains the full calculation.

The two terms in \eqref{eq:fixed-queue-order} arise from different mechanisms. The $1/k$ term is the residual batching/arrival contribution: increasing the batch size spreads the second-moment cost of an ordinary one-test cycle over more samples. The term $p_0k^2$ comes from positive pools. A positive pool occurs with probability approximately $kp_0$, but when it occurs the testing cycle contains $k$ additional individual tests; its contribution to the service-time second moment is therefore of order $p_0k^3$, which becomes $p_0k^2$ after division by batch size.

This is the key difference from classical Dorfman optimization. Expected
testing effort contains the retesting term $p_0k$, whereas queueing delay
contains the larger variability term $p_0k^2$. Consequently the same rare
positive pools enter the two objectives at different orders. Balancing
$1/k$ with $p_0k$ gives a square root; balancing $1/k$ with $p_0k^2$ gives a
cube root.

Lemma~\ref{lem:fixedBSQ} therefore already identifies the cube-root mechanism.
The next result obtains its exact constant from the mean waiting-time formula. The detailed proof is found in Appendix \ref{app:proof-exact-cuberoot-fixed}.

\begin{theorem}[Exact fixed-batch cube-root law]
\label{thm:exact-cuberoot-fixed}
Let $p_0\downarrow0$ and set $k\sim x p_0^{-1/3}$ with $x>0$.  Then
\begin{equation}
\E[W_{\rm queue}(k)]
\sim
\frac{\lambda}{2\mu^2}
\left(\frac2x+x^2\right)p_0^{1/3}.
\label{eq:exact-cuberoot-fixed}
\end{equation}
Moreover, within the cube-root family $k\sim x p_0^{-1/3}$, the leading
constant in \eqref{eq:exact-cuberoot-fixed} is uniquely minimized at $x=1$,
where it equals $3\lambda/(2\mu^2)$.
\end{theorem}
\paragraph{Proof sketch.} On the scale $k\sim xp_0^{-1/3}$, we have $p_k\sim kp_0$ and $k^3p_0\to x^3$. Substitution of the resulting testing-time moment expansions into the exact renewal-reward formula gives \eqref{eq:exact-cuberoot-fixed}. The excess-backlog term is only $O(p_0^{2/3})$ by \eqref{eq:fixed-backlog-lower-order}, and is therefore of lower order. The optimization of the resulting constant $2/x+x^2$ is immediate and gives the unique minimizer $x=1$. The detailed proof is found in Appendix \ref{app:proof-exact-cuberoot-fixed}.

The fixed-batch cube-root law gives the cleanest quantitative expression of the
variance mechanism. At the minimizing constant $x=1$, the two contributions
$2/x$ and $x^2$ in \eqref{eq:exact-cuberoot-fixed} are of the same order.
Moving to larger pools reduces the batching contribution but increases the
rare-positive service contribution quadratically. Moving to smaller pools does
the opposite. The cube root is therefore not an arbitrary interpolation
between individual testing and Dorfman pooling; it is the point at which these
two exact queueing effects balance.

It is also useful to compare the scale already at this stage with the
classical rule. Since $p_0^{-1/3}/p_0^{-1/2}=p_0^{1/6}\to0$, a
queueing-sensitive laboratory uses asymptotically much smaller pools than a
test-count optimizer. Section~\ref{sec:operational-objectives} will turn this
pointwise scaling comparison into a statement about global optimizers.

\subsection{State-dependent batches: exact cube-root asymptotics}
\label{subsec:exact-cuberoot-state}

The $(l,k)$ policy sharpens the interpretation of the cube-root regime. The
lower threshold $l$ determines when service normally starts; the upper
threshold $k$ matters only when enough backlog has accumulated to justify a
larger batch. The next results make this separation asymptotically exact.

\begin{lemma}[State-dependent queueing order]
\label{lem:statedepBSQ}
Let $p_0\downarrow0$, $k\to\infty$, $k^2p_0\to0$, and
$l/k\to\gamma\in(0,1]$.  Then
\begin{equation}
\E[W_{\rm queue}(l,k)]
=
\Theta\!\left(\frac1k+k^2p_0\right).
\label{eq:state-queue-order}
\end{equation}
\end{lemma}

\paragraph{Proof sketch.}
When $l/k\to\gamma>0$, every admissible batch size is uniformly of order $k$. Consequently the service-time second moment is uniformly of order $1+k^3p_0$, and the exact renewal-reward formula again produces the balance $1/k+k^2p_0$. The proof must additionally control the state-dependent backlog and the random batch size uniformly over $b\in\{l,\ldots,k\}$. Those details are provided in Appendix~\ref{app:proof-state-dependent-order}.

We now specialize to the cube-root scale.

\begin{proposition}[Collapse to the lower threshold]
\label{prop:state-collapse}
Let $p_0\downarrow0$,
$k\sim x p_0^{-1/3}$, and $l\sim y p_0^{-1/3}$, where
$0<y\leq x$.  Then,
\begin{displaymath}
    \Pr(B=l)=1-O(p_0^{1/3}),\qquad\E[B]\sim l.
\end{displaymath}
\end{proposition}

\paragraph{Proof sketch.}
On the cube-root scale the system is lightly loaded, so a new testing cycle almost always starts after an idle accumulation period rather than from a large backlog. The probability of beginning a cycle above the lower threshold therefore vanishes, implying that the selected batch size equals $l$ with probability tending to one. Appendix~\ref{app:proof-collapse-lower-threshold} gives the exact probability bounds.

\begin{theorem}[Exact state-dependent cube-root law]
\label{thm:exact-cuberoot-state}
Under the assumptions of Proposition~\ref{prop:state-collapse},
\begin{equation}
\E[W_{\rm queue}(l,k)]
\sim
\frac{\lambda}{2\mu^2}
\left(\frac2y+y^2\right)p_0^{1/3}.
\label{eq:exact-cuberoot-state}
\end{equation}
Within the family $l\sim y p_0^{-1/3}$, the leading constant is uniquely
minimized at $y=1$. If in addition $l/k\to\gamma\in(0,1]$, this minimizing
choice corresponds to $k\sim \gamma^{-1}p_0^{-1/3}$.
\end{theorem}
\paragraph{Proof sketch.} On the scale $l\sim yp_0^{-1/3}$, $k\sim xp_0^{-1/3}$, we have $p_b\sim bp_0$ and $b^3p_0=O(1)$, uniformly for $b\in\{l,\ldots,k\}$. Proposition \ref{prop:state-collapse} gives $\Pr(B=l)\to1$ and $\E[B]\sim l$. Substitution of these facts and the corresponding testing-time moment expansions into the exact renewal-reward formula gives \eqref{eq:exact-cuberoot-state}. All contributions from $B>l$ and the excess-backlog term are $O(p_0^{2/3})$ and therefore are of lower order. The optimization of $2/y+y^2$ gives the unique minimizer $y=1$. If $l/k\to\gamma$, then $k\sim\gamma^{-1}p_0^{-1/3}$. The proof is found in Appendix \ref{app:proof-exact-cuberoot-state}.

The state-dependent theorem does more than reproduce the fixed-batch exponent.
It also puts analytical structure around the two-threshold dynamic policy
reported numerically by \cite{benjaafar2025congestion}. On
the cube-root scale, service starts at $l$ with probability tending to one,
whereas $k$ is reached only under atypical backlog. The cube-root law therefore
describes the normal release threshold; the upper threshold remains a distinct
control for congestion. Fixed batching obscures this separation because it
forces the two decisions to coincide.

Operationally, this is the main role of the $(l,k)$ extension. The theorem
turns the cube-root law from a statement about a fixed pool size into a
statement about the \emph{normal release threshold}. With probability tending
to one, the laboratory starts with $l$ samples; the upper threshold $k$ is
relevant only in atypical backlog states. Thus the model separates two controls
that classical pooling treats as one.

The lower threshold can be chosen for responsiveness without giving up the
option to use larger batches when the queue is already long. In particular, for fixed
$l/k\to\gamma>0$, even the congestion cap remains on the cube-root scale and is
therefore asymptotically smaller than Dorfman's inverse-square-root
recommendation. The dynamic policy does not overturn the fixed-batch mechanism;
it decomposes it into a routine release decision and a rare-state congestion
decision.

Theorem \ref{thm:exact-cuberoot-state} is an asymptotic characterization within the scaling family $l\sim y p_0^{-1/3}$, $k\sim x p_0^{-1/3}$. It does not assert a globally optimal pair $(l,k)$ over all admissible thresholds. Rather, it identifies the leading-order behavior of the routine release threshold, while $k$ remains a separate congestion-control parameter.

\subsection{Balancing the three delay components}
The asymptotic results above deliberately isolate queueing because doing so
reveals the second-moment mechanism cleanly. A laboratory, however, observes
total turnaround rather than an abstract queueing component. It is therefore
useful to summarize how the three pieces move as the batch size changes before
turning to optimal pool sizing. Although the preceding analysis also considers the state-dependent $(l,k)$ policy, the numerical illustration in this subsection refers to the fixed-batch model $l=k$.

For $l=k$, every formed batch to be tested has size $k$, and the renewal-reward decomposition is $\bar W=W_{\rm idle}+W_{\rm queue}$, $W=W_{\rm idle}+W_{\rm queue}+W_{\rm test}$. The two pre-testing components $W_{\rm idle}$, $W_{\rm queue}$ have different operational interpretations.
The idle-time component $W_{\rm idle}$ is accumulated while the testing facility is available but the next batch is not yet complete. The busy-testing component $W_{\rm queue}$ is accumulated while another batch is being tested. It therefore includes both the waiting of an already formed batch and the part of the tagged batch's formation period during which the facility is occupied.

The fixed-batch formation benchmark $\frac{k-1}{2\lambda}$ is useful as a reference for batch formation, but it is not, in general, equal to $\E[W_{\rm idle}(k)]$. The former measures the average time required to collect the remaining samples needed to complete a batch, whereas the latter classifies the pre-testing delay according to whether the testing facility is idle.

Increasing $k$ therefore affects the three components in different ways. It changes the time spent completing a batch while the facility is idle, reduces the frequency of testing cycles, and may increase busy-testing delay because a positive pool generates a long and variable testing cycle. It may also increase the time required to classify a sample once testing has begun. The asymptotic results above isolate the busy-testing component, while the numerical results illustrate its tradeoff with the idle-time and in-testing components.

Table \ref{tab:delay-decomposition-renewal-reward} illustrates this tradeoff between the three components of turnaround time for the fixed-batch case $l=k$. The entries are normalized by the mean duration $1/\mu$ of one testing stage. We observe that increasing the pool size from $k=3$ to $k=7$ reduces the utilization from $0.693$ to $0.346$ and decreases the busy-testing component from $1.7385$ to $0.7181$. This reduction occurs at the same time with an increase in the idle-time component from $0.1968$ to $1.1655$, while the in-testing component also increases.

For comparison, the classical Dorfman testing criterion is $T(k)=\frac{1}{k}+p_k$, $p_k=1-(1-p_0)^k$, and for $p_0=0.0043385$ this criterion is minimized for $k=16$. On the other hand, the total turnaround time in Table \ref{tab:delay-decomposition-renewal-reward} is minimized at $k=4$. This difference illustrates that a pool size that is efficient in terms of the expected number of tests need not be optimal when turnaround time and testing congestion are taken into account.

\begin{table}[t]
\centering
\caption{\textbf{Turnaround-time decomposition for the renewal-reward fixed-batch case $l=k$, under normalized capacity.}
Exact mean delay components for $p_0=0.0043385$ and $\lambda/\mu=2$.
Time is normalized by the mean duration $1/\mu$ of one testing stage, so all the reported delay columns are $\mu\E[\cdot]$.}
\label{tab:delay-decomposition-renewal-reward}
\begin{tabular}{c c c c c c}
\toprule
$k$ & $\rho_k$ & $\mu\E[W_{\rm idle}]$ & $\mu\E[W_{\rm queue}]$
& $\mu\E[W_{\rm test}]$ & $\mu\E[W]$ \\
\midrule
3  & 0.693 & 0.1968   &  1.7385  &   1.0259 & 2.9612 \\
4  & 0.534 & 0.4351   &  0.9718  &   1.0431 & 2.4501 \\
7  & 0.346 & 1.1655   &  0.7181   &  1.1199 & 3.0036 \\
16 & 0.259 & 3.0999   &  1.6652  &   1.5712 & 6.3363 \\
\bottomrule
\end{tabular}
\end{table}

\section{\texorpdfstring{When does congestion modify Dorfman's square-root rule?}{When does congestion modify Dorfman's square-root rule?}}
\label{sec:extendDorfman}

The previous section shows that queueing delay can select a cube-root pool-size
scale even when the testing-effort optimum follows Dorfman's square-root rule.
There is, however, a different question that is equally important for
capacity-constrained laboratories: \emph{can congestion change the classical
Dorfman recommendation while the pool size itself remains on the square-root
scale?} The answer is yes, but only under a particular joint scaling of
prevalence and testing capacity.

This section isolates that regime. The distinction is useful because it
separates two ways in which operations can modify classical pooling. Under
fixed capacity and a delay-sensitive objective, the relevant exponent can
change from $1/2$ to $1/3$, as established in Section~\ref{sec:exact-asymptotics}
and formalized globally in Section~\ref{sec:operational-objectives}. By
contrast, when testing capacity itself becomes scarce as prevalence decreases,
congestion can enter at first order \emph{within} the classical square-root
domain. In that case the exponent survives but the optimal constant changes.

\subsection{Why fixed capacity does not create a congestion correction}
Throughout this subsection, $\lambda$ and $\mu$ are fixed positive constants, and $p_0\downarrow0$. The classical Dorfman benchmark minimizes expected tests per sample,
$T(k)=1/k+p_k$, and satisfies $k_D^\star\sim p_0^{-1/2}$. To see whether
ordinary queueing congestion changes this recommendation, take
$k=x/\sqrt{p_0}$ with $x>0$. Then $p_k\sim x\sqrt{p_0}$ and
$kp_k\sim x^2$, so $\E[S^{(k)}]\sim\frac{1+x^2}{\mu}$.
Consequently, 
\[
\rho_k=
\frac{\lambda\E[S^{(k)}]}{k}
=
\Theta(\sqrt{p_0})
\longrightarrow0.
\]

Thus the testing facility becomes asymptotically empty at the classical
Dorfman scale. This observation is important because it rules out a common
intuition: with fixed capacity, the square-root rule does \emph{not} remain
interesting because the laboratory is increasingly congested. The opposite
occurs. Pooling reduces the number of testing cycles so aggressively that
utilization vanishes.

The renewal-reward delay consequence is also immediate. In particular, at the square-root scale
\begin{displaymath}
    \E[(S^{(k)})^2]=\frac{2(1-p_k)+p_k(k+1)(k+2)}{\mu^2}\sim \frac{x^3}{\mu^2 p_0^{1/2}}=\frac{x^3}{\mu^2}\frac{k}{x}=\frac{x^3}{\mu^2}k=\Theta(k).
\end{displaymath}
Moreover, if $A^{(k)}$ be the number of arrivals during one testing cycle, then
\begin{displaymath}
    {\rm Var}(A^{(k)})=\lambda\E[S^{(k)}]+\lambda^2 {\rm Var}(S^{(k)})=O(k),
\end{displaymath}
whereas $k-\E[A^{(k)}]=k-\lambda\E[S^{(k)}]=\Theta(k)$. Moreover, the Kingman bound for the associated Lindley recursion therefore gives
\begin{displaymath}
    \E[(Q-k)^{+}]\leq \frac{ {\rm Var}(A^{(k)})}{2(k-\E[A^{(k)}])}=\frac{O(k)}{\Theta(k)}=O(1).
\end{displaymath}
Therefore, 
\begin{displaymath}
    \frac{\E[S^{(k)}]}{k}\E[(Q-k)^+]=O(k^{-1}).
\end{displaymath}
By substituting to the exact renewal-reward representation we get
\begin{displaymath}
    \E[W_{\rm queue}(k)]=\frac{\lambda x^2}{2\mu^2}+o(1)=\Theta(1).
\end{displaymath}
Thus, the busy-testing component is of order $O(1)$. It remains to identify the order of the idle-time component. Recall that $W_{\rm form}$ denotes the formation-delay benchmark. Note also that $W_{\rm idle}\leq W_{\rm form}$, because $W_{\rm idle}$ is only the portion of the formation interval during which the testing facility is idle. Then, $W_{\rm form}\leq \bar W =W_{\rm idle}+W_{\rm queue}$, since a tagged batch must be first formed before testing begins. Therefore,
\begin{displaymath}
    \E[W_{\rm form}]-\E[W_{\rm queue}(k)]\leq \E[W_{\rm idle}(k)]\leq \E[W_{\rm form}].
\end{displaymath}
Since $\E[W_{\rm form}]=\frac{k-1}{2\lambda}$,
\begin{displaymath}
    0\leq \frac{k-1}{2\lambda}-\E[W_{\rm idle}]\leq \E[W_{\rm queue}(k)]= O(1),
\end{displaymath}
thus,
\begin{displaymath}
    \E[W_{\rm idle}(k)]=\frac{k-1}{2\lambda}+O(1)=\Theta(k)=\Theta(p_0^{-1/2}).
\end{displaymath}
Finally, $\E[W_{\rm test}(k)]=\frac1\mu+\frac{p_k(k+1)}{2\mu}\to \frac{1+x^2/2}{\mu}=O(1)$. Consequently, the renewal-reward total turnaround time satisfies
\begin{displaymath}
    \E[W]=\frac{k-1}{2\lambda}+O(1)=\Theta(p_0^{-1/2}).
\end{displaymath}
Thus, within the classical square-root rule, fixed-capacity congestion does not generate a first-order correction to the Dorfman scale. The dominant contribution is idle-time batch formation, while both the busy-testing and in-testing components remain bounded. Note that the benchmark $W_{\rm form}$ is used here only to obtain the order of $W_{\rm idle}$.


\subsection{The critical capacity scaling}

For congestion to remain visible while $k=\Theta(p_0^{-1/2})$, testing
capacity must deteriorate jointly with prevalence. Since
$\E[S^{(k)}]=\Theta(1/\mu)$ on this scale, $\rho_k
=
\Theta\!\left(
\frac{\lambda}{\mu}\sqrt{p_0}
\right)$.
Therefore utilization remains of order one if and only if
\begin{equation}
\frac{\mu}{\lambda}
=
\Theta(\sqrt{p_0}).
\label{eq:critical-capacity-scaling}
\end{equation}

This is the unique congestion-preserving order at the Dorfman scale. If
$\mu/\lambda$ decreases more slowly than $\sqrt{p_0}$, utilization tends to
zero and congestion disappears. If it decreases more rapidly, the offered load
eventually exceeds capacity and the square-root family becomes unstable. Thus
\eqref{eq:critical-capacity-scaling} is not an arbitrary heavy-traffic
parameterization: it is the boundary between an asymptotically slack testing
facility and one that cannot sustain Dorfman-scale pools.

A useful way to interpret the scaling is to view prevalence and capacity as
joint determinants of the effective workload. Low prevalence favors larger
pools and therefore fewer pooled tests per incoming sample. Scarce testing
capacity pushes in the opposite direction. Equation
\eqref{eq:critical-capacity-scaling} identifies the rate at which these effects
exactly offset one another.

\subsection{Scarce capacity and the exact square-root correction}
In this subsection, we assume the arrival rate $\lambda>0$ fixed, and for $\varepsilon\downarrow0$ let $p_0=\varepsilon^2$, $k=\frac{x}{\varepsilon}$, and the testing rate to decrease according to $\frac{\mu}{\lambda}=c\varepsilon$, $c>2$.
Then $p_k\sim x\varepsilon$ and $p_k+\frac1k
\sim
\varepsilon\left(x+\frac1x\right).$
Moreover,
$\E[S^{(k)}]\sim(1+x^2)/\mu$, and thus, $\rho_k\longrightarrow
\rho(x)=\frac{x+1/x}{c}$.
The limiting stability condition is therefore $x+1/x<c$. Since
$x+1/x\ge2$, a feasible interval exists only if $c>2$. In that case
$x_-<x<x_+$, where $x_\pm=\frac{c\pm\sqrt{c^2-4}}{2}$.

The limiting utilization already shows that scarce capacity changes the
feasible set of Dorfman-scale pools. The more subtle effect comes from
variability. On the same scale, $c_{S}^2(k)
\sim
\frac{x^3}{(1+x^2)^2}\frac1{\varepsilon}$.
Thus the service-time coefficient of variation diverges even though positive
pools are rare. The reason is the same rare-event mechanism encountered in
Section~\ref{sec:exact-asymptotics}: a positive pool occurs with probability of
order $\varepsilon$, but generates a burst of order $\varepsilon^{-1}$
additional tests. Near capacity, this variability is amplified strongly enough
to determine the leading turnaround time.

The scarce-capacity analysis below concerns the total turnaround time $\E[W(k)]=\E[W_{\rm idle}(k)]
+\E[W_{\rm queue}(k)]+\tau_k$, where the pre-testing delay $\E[\bar W(k)]=\E[W_{\rm idle}(k)]+\E[W_{\rm queue}(k)]$ is evaluated through the exact root representation. The conventional formation-delay benchmark $W_{\rm form}$ is not used in this analysis.

\begin{theorem}[Exact congestion-corrected Dorfman constant]
\label{thm:extendDorfman}
Let $p_0=\varepsilon^2\downarrow0$, let
$k\sim x/\varepsilon$, and suppose $\mu/\lambda=c\varepsilon$ with $\lambda>0$ and $c>2$ be fixed. Define $x_\pm=\frac{c\pm\sqrt{c^2-4}}{2}$, $x\in(x_-,x_+)$. Then, for the fixed-batch Dorfman system,
\begin{equation}
\E[W(k)]
\sim
\frac{1}{2c\lambda\varepsilon^2}
\frac{x^3}{cx-x^2-1}.
\label{eq:exact-scarce-turnaround}
\end{equation}
Consequently, within the square-root family \(k\sim x/\varepsilon\) we have
\begin{equation}
k^\star
\sim
\frac{x^\star(c)}{\sqrt{p_0}},
\qquad
x^\star(c)=c-\sqrt{c^2-3}.
\label{eq:congestion-corrected-constant}
\end{equation}
\end{theorem}

\paragraph{Proof sketch.}
Under the critical capacity scaling, $k\sim x p_0^{-1/2}$ keeps the Dorfman workload and congestion on the same first-order scale. Expanding the exact fixed-batch quantities on this scale gives a one-dimensional limiting objective in $x$. Its first-order condition identifies the capacity-dependent constant, and the boundary behavior rules out escape from the interior. The full expansions and optimization argument are in Appendix~\ref{app:proof-congestion-corrected-dorfman}.

Theorem \ref{thm:extendDorfman} identifies a precise sense in which the classical square-root rule
is robust. Congestion does not necessarily change the exponent. Under the
critical capacity scaling it changes the \emph{constant}:
$k^\star=x^\star(c)p_0^{-1/2}[1+o(1)]$. The parameter $c$ summarizes how much
testing capacity remains relative to incoming demand at the low-prevalence
scale. As $c$ decreases toward the critical value $2$, the feasible interval narrows, but the turnaround-optimal constant $x^\star(c)$ increases toward the classical Dorfman constant $1$. In contrast, as $c$ increases, $x^\star(c)$ decreases and the deviation from the classical constant becomes larger.

The correction also has a direct comparison with classical Dorfman. The
classical leading constant is one, whereas the operational constant is
$x^\star(c)$. Thus the cost of ignoring congestion is first-order whenever
$c$ is finite. What remains unchanged is the $p_0^{-1/2}$ rate itself. This is
precisely the kind of correction one would expect a manager to use when
statistical test efficiency remains the relevant scale but capacity is not
abundant.

It is useful to contrast this result with the cube-root law. The two results do
not describe competing approximations to the same regime. The
congestion-corrected square root arises because capacity is itself scaled so
that utilization stays nondegenerate at $k=\Theta(p_0^{-1/2})$. The cube-root
law arises under fixed $\lambda,\mu$ when the objective values the queueing
consequences of service-time variability. One changes the leading constant
within the classical domain; the other changes the relevant pool-size scale.

\subsection{Where the square-root regime sits in the broader asymptotic map}

The preceding theorem is the only nondegenerate congestion correction needed
for our main square-root analysis. It is nevertheless helpful to locate it
relative to neighboring scales.

For pool sizes satisfying $k^2p_0\to0$, positive pools are so rare that
$\E[S^{(k)}]\sim1/\mu$ and $\rho_k\sim\lambda/(\mu k)$. The testing cycle is
essentially a single test. Increasing $k$ then has little effect on congestion,
while formation delay continues to grow. This is the low-prevalence region in
which the cube-root queueing balance of Section~\ref{sec:exact-asymptotics}
emerges.

At the Dorfman scale $k=\Theta(p_0^{-1/2})$, the expected numbers of pooled and
follow-up tests are comparable. With fixed capacity, utilization still tends
to zero; with the critical scaling \eqref{eq:critical-capacity-scaling},
congestion remains first-order and Theorem~\ref{thm:extendDorfman} applies.

For much larger pools, for example $k=\Theta(p_0^{-1})$, the probability of a
positive pool no longer vanishes. The rare-positive mechanism underlying both
the classical Dorfman expansion and the cube-root queueing law has then
disappeared. Such scales can be useful as boundary cases, but they no longer
describe low-prevalence pooling in the classical sense and do not generate a
new optimality result in the present paper.

Heavy traffic is likewise best understood through the exact result already
proved in Proposition~\ref{prop:exact-heavy-traffic}. When $\rho_k\uparrow1$,
the factor $(1-\rho_k)^{-1}$ amplifies testing-time variability. The
scarce-capacity regime above is the square-root-scale instance in which that
amplification remains first-order without destroying stability.

\begin{table}[t]
\centering
\caption{\textbf{Capacity correction to the Dorfman square-root scale.}
Under $p_0=\varepsilon^2$ and $\mu/\lambda=c\varepsilon$, a policy
$k\sim x/\sqrt{p_0}$ is stable for $x\in(x_-(c),x_+(c))$, where
$x_\pm(c)=(c\pm\sqrt{c^2-4})/2$. Theorem~\ref{thm:extendDorfman} gives the
turnaround-optimal constant $x^\star(c)=c-\sqrt{c^2-3}$.}
\label{tab:scarce-capacity}
\begin{tabular}{c c c c}
\toprule
$c$ & $x_-(c)$ & $x^\star(c)$ & $x_+(c)$ \\
\midrule
2.05 & 0.800 & 0.953 & 1.250 \\
2.25 & 0.610 & 0.814 & 1.640 \\
2.50 & 0.500 & 0.697 & 2.000 \\
3.00 & 0.382 & 0.551 & 2.618 \\
4.00 & 0.268 & 0.394 & 3.732 \\
5.00 & 0.209 & 0.310 & 4.791 \\
\bottomrule
\end{tabular}
\end{table}

Table~\ref{tab:scarce-capacity} reports the closed-form quantities from
Theorem~\ref{thm:extendDorfman}. When $c$ is close to the critical value $2$,
the stability interval is narrow, so capacity strongly restricts which
square-root constants are feasible. As $c$ increases, the interval widens, while both $x_-(c)$ and $x^\star(c)$ decrease. Moreover, as $c$ increases, the turnaround-optimal constant $x^\star(c)$ approaches the lower stability boundary $x_-(c)$. The key point is that congestion does not create a new
exponent in this regime: the optimal pool remains of order $p_0^{-1/2}$, but
its leading constant is determined by the capacity parameter $c$.

The message of this section is intentionally narrow. It answers when congestion
modifies the classical square-root rule \emph{without changing its exponent}.
The next section asks a different design question: once the laboratory changes
the objective itself---from tests alone to queueing, combined costs, or
formation-sensitive performance---which asymptotic scale does the global
optimizer select? Keeping these questions separate prevents the
congestion-corrected square root from being confused with the cube-root
delay-sensitive regime.

\section{Operational objectives and optimal scaling}
\label{sec:operational-objectives}

The previous sections characterize how pooling changes the stochastic dynamics
of the testing system. We now turn from performance analysis to the design
question that is ultimately relevant for practice: \emph{which pool size should
a laboratory choose?} The answer depends on what the laboratory is trying to
optimize. Classical Dorfman pooling minimizes the expected number of tests.
\cite{benjaafar2025congestion} show numerically that replacing
this throughput-oriented criterion by expected delay can materially change the
recommended pool size. We take the next step and ask which asymptotic scale is
selected by each objective. An operational system may care about queueing
delay, or may attach explicit costs to testing effort, queueing, and pool
formation. Once these objectives are distinguished, the different scaling laws
derived earlier acquire a direct decision-theoretic interpretation.

This section is therefore not another asymptotic approximation of the queue. Its purpose is to convert the exact results of Sections~\ref{sec:exact-asymptotics}--\ref{sec:extendDorfman} into statements about \emph{global integer optimizers}. This distinction is important. Balancing two leading-order terms can suggest a candidate scale, but does not by itself exclude much smaller or much larger pool sizes. The arguments below use the exact queueing bounds to rule out such alternatives.

Throughout, $p_0\downarrow0$ while $\lambda$ and $\mu$ remain fixed. For each $p_0$, optimization is over stable integer pool sizes. We write
$T(k)=1/k+p_k$ for expected tests per sample, where $p_k=1-(1-p_0)^k$. Whenever $kp_0\to0$, the familiar expansion $p_k\sim kp_0$ applies. Here we have to distinguish the delay quantities according to the objective under consideration. The queueing results in Sections \ref{sec:lower-bound}-\ref{sec:joint-testing} refer to the renewal-reward busy-testing component $W_{\rm queue}$ defined in \eqref{eq:delay-comp}. The turnaround-time decomposition is $W=W_{\rm idle}+W_{\rm queue}+W_{\rm test}$.

On the other hand, Section \ref{sec:formation-benchmark} considers another formation-sensitive objective based on the batch-formation benchmark $\E[W_{\rm form}(k)]=\frac{k-1}{2\lambda}$. As said above, this benchmark is not identical to the renewal-reward idle component $W_{\rm idle}$, and it is not an additional term in the turnaround-time decomposition. It is used only to study objectives that assign a positive cost to conventional batch formation.


The three relevant balances can already be seen from these expressions and from the exact queueing results obtained earlier. Test efficiency balances the pooled-test term $1/k$ against the expected retesting term $kp_0$, suggesting the classical square-root scale. Queueing delay instead contains the rare-event variability term $p_0k^2$; Lemma~\ref{lem:fixedBSQ} shows rigorously that, under $k^2p_0\to0$, $\E[W_{\rm queue}(k)]
=\Theta\!\left(\frac1k+p_0k^2\right)$, which suggests the cube-root scale. The formation-delay benchmark behaves differently: it grows linearly in $k$ and does not vanish as prevalence decreases. This will force the optimizer to remain bounded whenever formation carries a positive cost.

These are not competing approximations to the same optimization problem. They arise because different objectives put weight on different consequences of pooling. This also formalizes a tradeoff noted in prior operational work: test cost and delay need not favor the same pool size \cite{benjaafar2025congestion,bish2021robust}. The results below make that
statement precise at the level of global optimizer scaling.

\subsection{A global lower bound for queueing delay}\label{sec:lower-bound}

Before optimizing the queueing-based objectives, we need one ingredient that is stronger than the local asymptotic formula in Lemma~\ref{lem:fixedBSQ}. That lemma describes the queue once $k$ is already in the low-prevalence range $k^2p_0\to0$. For a global optimization result, however, we must first show that an optimizer cannot escape that range. See Appendix~\ref{app:proof-global-queue-lower} for the proof.

\begin{lemma}[Global queueing lower bound]
\label{lem:global-queue-lower}
For every stable fixed-batch Dorfman system,
\begin{equation}
\E[W_{\rm queue}(k)]
\ge
\frac{\lambda}{2k}\E[(S^{(k)})^2].
\label{eq:queue-global-lower}
\end{equation}
Moreover, let $p_0\downarrow0$ along any sequence of stable pool sizes for which $\E[W_{\rm queue}(k)]\to0$. Then necessarily $k\to\infty$ and $kp_0\to0$. Along every such sequence there exist constants $a,b>0$, independent of $p_0$ and $k$, such that for all sufficiently small $p_0$,
\begin{equation}
\E[W_{\rm queue}(k)]
\ge
\frac{a}{k}+b p_0k^2.
\label{eq:queue-global-two-sided-lower}
\end{equation}
\end{lemma}

\paragraph{Proof sketch.}
Dropping the nonnegative excess-backlog term from the exact renewal-reward formula gives the first lower bound immediately. If queueing delay vanishes, that bound first forces $k\to\infty$ and then $kp_0\to0$. In this region $p_k\sim kp_0$, so the same exact second-moment expression yields simultaneous lower bounds of order $1/k$ and $p_0k^2$. See Appendix~\ref{app:proof-global-queue-lower} for the complete contradiction arguments and constants.

Lemma~\ref{lem:global-queue-lower} provides the compactness argument that the heuristic balance alone is missing. A candidate pool size of cube-root order makes queueing delay vanish at rate $p_0^{1/3}$; the lemma then forces any global optimizer with comparable performance into the same asymptotic region.

\subsection{Classical benchmark: minimizing the number of tests}\label{sec:classic-benchmark}

We begin with the criterion for which Dorfman's rule was originally designed. This benchmark is useful not only historically but also as a reference against which the operational objectives can be measured.

\begin{theorem}[Classical Dorfman benchmark]
\label{thm:global-dorfman}
Let $k_T^\star$ minimize $T(k)=1/k+p_k$ over stable integer pool sizes. Then
\begin{equation}
k_T^\star\sim p_0^{-1/2}.
\label{eq:domain-testing}
\end{equation}
\end{theorem}

\paragraph{Proof sketch.}
The candidate $k\asymp p_0^{-1/2}$ attains testing effort of order $\sqrt{p_0}$ and is eventually stable. Optimality then forces both $k\to\infty$ and $kp_0\to0$, reducing the objective to $1/k+kp_0$ to first order. Rescaling by $x=k\sqrt{p_0}$ leaves the uniquely minimized function $x+1/x$. The full integer-optimization argument is given in Appendix~\ref{app:proof-global-dorfman}.

Theorem~\ref{thm:global-dorfman} recovers the classical square-root rule as a global integer optimization result. Its economic logic is simple: increasing $k$ reduces the pooled-test burden $1/k$, but raises the expected number of individual retests by approximately $kp_0$. Balancing these two \emph{mean testing-effort} terms yields $k\asymp p_0^{-1/2}$.

The next result shows why this rule can be far too aggressive once the objective changes from test counts to delay.

\subsection{Queueing delay and the cube-root rule}\label{sec:queueing-delay-cube-root}

Queueing is sensitive not only to the mean number of tests but also to the variability of the testing cycle. A positive pool is rare when $p_0$ is small, yet it creates a burst of $k$ follow-up tests. The second moment therefore contains a term of order $p_0k^3$, which becomes $p_0k^2$ after normalization by batch size. This is the source of the cube-root balance.

\begin{theorem}[Global queueing-optimal scale]
\label{thm:global-queue-cuberoot}
Let $k_Q^\star$ minimize the exact fixed-batch queueing delay $\E[W_{\rm queue}(k)]$ over stable integer pool sizes. Then, 
\begin{equation}
k_Q^\star=\Theta(p_0^{-1/3}).
\label{eq:domain-queueing}
\end{equation}
More precisely, 
\begin{displaymath}
    k_Q^\star\sim p_{0}^{-1/3}.
\end{displaymath}
\end{theorem}

\paragraph{Proof sketch.}
A cube-root candidate makes queueing delay $O(p_0^{1/3})$. The global lower bound then prevents an optimizer from using either a smaller or a larger asymptotic scale, giving $k_Q^\star=\Theta(p_0^{-1/3})$. To identify the constant, rescale $k_Q^\star p_0^{1/3}$, take an arbitrary convergent subsequence, and apply the exact cube-root limit; uniqueness of the minimizer of its limiting objective forces every accumulation point to equal one. The full subsequence argument is in Appendix~\ref{app:proof-global-queue-cuberoot}.

The conclusion differs sharply from Dorfman. Since $k_T^\star\sim p_0^{-1/2}$, we have that $\frac{k_Q^\star}{k_T^\star}=\Theta(p_0^{1/6})\longrightarrow0$. Thus, the gap between the test-minimizing and queueing-minimizing prescriptions becomes asymptotically large at low prevalence. Classical pooling keeps increasing the group size because tests become statistically cheap to save; the queueing objective stops much earlier because the workload burst generated by a rare positive pool grows with $k$.

The second assertion in Theorem \ref{thm:global-queue-cuberoot} substantially sharpens the order result in \eqref{eq:domain-queueing} by identifying the limiting constant. In particular,
\begin{displaymath}
    k_Q^\star \sim p_0^{-1/3},
\end{displaymath}
rather than merely $k_Q^\star=\Theta(p_0^{-1/3})$. Table \ref{tab:consequences-queueing-optimizers} numerically justifies this asymptotic result.

This queueing result also explains the role of the dynamic $(l,k)$ policy introduced earlier. Theorem~\ref{thm:exact-cuberoot-state} shows that, on the same cube-root scale, the state-dependent system selects the lower release threshold $l$ with probability tending to one. Operationally, the cube-root rule therefore governs \emph{routine release}, while the upper threshold $k$ remains available as a separate congestion-control lever.

\subsection{Joint testing and queueing costs}\label{sec:joint-testing}

Most laboratories do not minimize queueing time in isolation. Reagents, machine cycles, labor, and delay can all matter. Let $c_T>0$ denote the cost assigned to one test and $c_Q>0$ the cost assigned to one unit of queueing time, and define
\[
J_{TQ}(k)=c_TT(k)+c_Q\E[W_{\rm queue}(k)].
\]
The weights convert the two quantities to a common cost scale. The next theorem shows that any fixed positive queueing weight is enough to change the asymptotic exponent from Dorfman's square root to a cube root.

\begin{theorem}[Testing plus queueing]
\label{thm:global-testing-queueing}
Let $k_{TQ}^\star$ minimize $J_{TQ}(k)$ over stable integer pool sizes. For fixed $c_T,c_Q>0$,
\begin{equation}
k_{TQ}^\star=\Theta(p_0^{-1/3}).
\label{eq:domain-combined}
\end{equation}
\end{theorem}

\paragraph{Proof sketch.}
At a cube-root candidate both testing effort and queueing delay are $O(p_0^{1/3})$. Since the two cost components are nonnegative, optimality forces the optimizer's queueing delay to vanish at that rate as well. The global queueing lower bound can therefore be applied directly, and its $1/k$ and $p_0k^2$ terms pin the optimizer to the cube-root scale. Full details appear in Appendix~\ref{app:proof-testing-queueing}.

The reason the testing term does not restore the square-root rule is revealing. At $k\asymp p_0^{-1/3}$, the pooled-test contribution $1/k$ is already of order $p_0^{1/3}$, whereas the retesting contribution $kp_0$ is only of order $p_0^{2/3}$. The leading testing cost therefore reinforces the $1/k$ side of the queueing balance rather than recreating the classical $1/k$ versus $kp_0$ tradeoff.

The exact cube-root asymptotics allow us to go one step further and identify how the cost weights move the leading constant within the cube-root family. The detailed proof is found in Appendix \ref{app:proof-weighted-cuberoot-constant}.

\begin{proposition}[Cost-dependent constant within the cube-root family]
\label{prop:weighted-cuberoot-constant}
Suppose $k\sim xp_0^{-1/3}$ with $x>0$. Then
\[
J_{TQ}(k)
\sim
\left[
\frac{c_T}{x}
+
\frac{c_Q\lambda}{2\mu^2}
\left(\frac{2}{x}+x^2\right)
\right]p_0^{1/3}.
\]
The leading constant is uniquely minimized at
\[
x_{TQ}^\star
=
\left(1+\frac{c_T\mu^2}{c_Q\lambda}\right)^{1/3}.
\]
\end{proposition}
\paragraph{Proof sketch.} On the scale $k\sim xp_0^{-1/3}$, we have $p_k\sim kp_0\sim xp_0^{2/3}$, and hence $T(k)=\frac{1}{k}+p_k\sim \frac{1}{x}p_0^{1/3}$. Combining with Theorem \ref{thm:exact-cuberoot-fixed} in the definition of $J_{TQ}(k)$ gives $J_{TQ}(k) \sim [\frac{c_T}{x} + \frac{c_Q\lambda}{2\mu^2} \left(\frac2x+x^2\right)]p_0^{1/3}$. This proves the stated asymptotic relation. Differentiating the coefficient with respect to $x$ gives the displayed value of $x_{TQ}^{\star}$. Its strict convexity on $(0,\infty)$ justifies its uniqueness. The detailed proof is found in Appendix \ref{app:proof-weighted-cuberoot-constant}.

This proposition makes the managerial effect of the weights transparent. A larger relative cost of tests increases the preferred cube-root constant and therefore favors larger pools; a larger cost of queueing pushes the constant back toward the queueing-only value $x=1$. The exponent, however, remains $1/3$ as long as both weights stay fixed and strictly positive.

\subsection{Formation delay benchmark and the bounded regime}\label{sec:formation-benchmark}

Formation delay benchmark changes the problem qualitatively. Unlike the retesting and queueing terms, the time required to assemble a fixed batch does not disappear as prevalence falls. For a Poisson arrival stream it equals $(k-1)/(2\lambda)$ and grows linearly in $k$. Consequently, once formation has any fixed positive cost, arbitrarily large pools cannot remain optimal.

Because optimization is restricted to stable policies, the benchmark used in this argument must itself be stable. Let $k_0$ be any fixed integer satisfying $k_0>\lambda/\mu$. Since $p_{k_0}\to0$, this $k_0$ is stable for all sufficiently small $p_0$. Here $W_{\rm form}$ denotes the formation delay benchmark, rather than the renewal-reward component $W_{\rm idle}$ defined in Section \ref{sec:model}. Thus, the bounded-pooling result below is stated for an objective that assigns a positive cost to this conventional formation benchmark (see Appendix \ref{app:proof-bounded-formation} for the proof).

\begin{theorem}[Positive formation-benchmark cost implies bounded pooling]
\label{thm:bounded-formation}
Let
\[
J_F(k)=c_F\E[W_{\rm form}(k)]+R_{p_0}(k),
\qquad c_F>0,
\]
where $R_{p_0}(k)\ge0$. Assume that for some fixed integer $k_0>\lambda/\mu$,
$\sup_{p_0\le p^\dagger}R_{p_0}(k_0)<\infty$ for a sufficiently small $p^\dagger>0$. If $k_F^\star$ minimizes $J_F$ over stable integer pool sizes, then as $p_0\downarrow0$, $k_F^\star=\Theta(1)$.
\end{theorem}

The theorem is intentionally general: $R_{p_0}$ may contain testing costs, queueing costs, in-testing delay, or other nonnegative operational penalties. The only requirement is that at least one fixed stable policy has uniformly bounded residual cost. The conclusion is correspondingly strong: a positive formation cost changes the \emph{exponent itself} to zero.

Note that Theorem \ref{thm:bounded-formation} refers to an objective function that contains the fixed batch formation benchmark. We now connect this benchmark to the renewal-reward total turnaround objective. Clearly, the benchmark is not identical to the renewal-reward idle component $\E[W_{\rm idle}(k)]$. However, it provides a lower bound for the total pre-testing delay, since a sample must first wait until its batch is formed. The next result establishes the asymptotic order of the total-turnaround optimizer $k_W^\star$.

\begin{corollary}[Bounded pool size under total-turnaround minimization]\label{cor:bounded-turnaround}
Fix $\lambda$ and $\mu$, and let $k_W^\star(p_0)$ minimize the exact total turnaround time $\E[W(k)]=\E[W_{\rm idle}(k)]+\E[W_{\rm queue}(k)]+\E[W_{\rm test}(k)]$, for all stable pool sizes $k$. Then, as $p_0\downarrow0$, $k_W^\star(p_0)=\Theta(1).$
\end{corollary}
\paragraph{Proof sketch.} Choose a fixed integer $k>\lambda/\mu$. Since $p_{k}\to0$, the policy with batch size $k$ is stable for all sufficiently small $p_0$, and its total turnaround time remains bounded. Hence optimality gives $\E[W(k_W^\star(p_0))]=O(1)$. On the other hand, every tagged sample must wait until its batch is formed, so $\E[W(k)]\geq \E[W_{\rm form}(k)] =\frac{k-1}{2\lambda}$. Thus, $k_W^\star(p_0)=O(1)$. Since $k_W^\star(p_0)\geq1$, we obtain $k_W^\star(p_0)=\Theta(1)$. The detailed proof is found in Appendix \ref{app:proof-bounded-turnaround}.

The preceding result refers to the renewal-reward total-turnaround objective and establishes only the order of its optimizer. It does not identify the limiting integer selected by that objective. We now return to the conventional testing-versus-formation criterion, for which a more precise limiting optimization problem can be obtained. For the numerical analysis it is useful to specialize this result to the simplest testing-versus-formation objective,
\[
J_{TF}(k)=c_TT(k)+c_F\E[W_{\rm form}(k)].
\]
Here we can say more than boundedness. The proof of the following result is found in Appendix \ref{app:proof-formation-limit} 

\begin{proposition}[Limiting pool size with testing and formation costs]
\label{prop:formation-limit}
Let $k_{TF}^\star$ minimize $J_{TF}$ over stable integer pool sizes, and define the limiting feasible set $\mathcal K_0=\{k\in\mathbb N:k>\lambda/\mu\}$. Then every accumulation point of $k_{TF}^\star$ belongs to $\arg\min_{k\in\mathcal K_0}
\left\{
\frac{c_T}{k}+\frac{c_F}{2\lambda}(k-1)
\right\}$. If this limiting discrete problem has a unique minimizer $k_{TF}^{\infty}$, then $k_{TF}^\star=k_{TF}^{\infty}$ for all sufficiently small $p_0$.
\end{proposition}

Ignoring integer and stability constraints for a moment, the limiting continuous objective is minimized at $k=\sqrt{\frac{2\lambda c_T}{c_F}}$. The operational prescription is therefore especially simple: choose an integer near this value, but not below the limiting stability requirement $k>\lambda/\mu$. This clarifies why the bounded regime may appear numerically as a flat curve. The plateau is not an artifact of the computation; it is the finite-prevalence manifestation of a genuine limiting optimization problem. If the stability floor binds, capacity rather than prevalence determines the eventual pool size.

\subsection{Comparison of the operational regimes}

The results can now be read as a hierarchy of increasingly operational objectives. Under test-count minimization, the classical Dorfman prescription is $k_T^\star\sim p_0^{-1/2}$. Under queueing minimization, the globally optimal pool size is only of order $p_0^{-1/3}$. Adding any fixed positive test cost to the queueing objective changes the leading constant but not this cube-root exponent. Adding a positive formation cost changes the behavior more fundamentally: the optimal pool size remains bounded.

Relative to Dorfman, both operationally delay-sensitive regimes therefore prescribe much smaller pools at low prevalence. In particular,
\[
\frac{k_Q^\star}{k_T^\star}=\Theta(p_0^{1/6})\to0,
\qquad
\frac{k_{TQ}^\star}{k_T^\star}=\Theta(p_0^{1/6})\to0,
\]
while any bounded formation-sensitive optimizer satisfies $k_F^\star/k_T^\star\to0$. These ratios give a direct interpretation to the scaling laws: the difference between classical and operational pooling is not a small correction that disappears at low prevalence. It becomes more pronounced.

For the dynamic $(l,k)$ policy, the same comparison should be interpreted through the release threshold. The state-dependent cube-root theorem in Section~\ref{subsec:exact-cuberoot-state} shows that $l$, rather than the congestion cap $k$, governs routine operation on the cube-root scale. Hence the operational lesson is not simply ``use a smaller fixed pool.'' It is to release normal batches on the delay-sensitive scale while retaining a separate upper threshold for absorbing backlog.

\begin{table}[t]
\centering
\caption{\textbf{Finite-prevalence consequences of the operational scaling
laws for the fixed-batch case $l=k$.} Numerically computed integer minimizers under the normalized capacity ratio
$\lambda/\mu=2$. The first column uses the \cite{Cabrera2023} prevalence and then lowers
prevalence successively to illustrate the asymptotic regimes.}
\label{tab:finite-prevalence-renewal}
\begin{tabular}{c c c c}
\toprule
Prevalence $p_0$ & Tests only ($k^{\star}_{T}$) & Queueing only ($k^{\star}_{Q}$) & Total turnaround ($k^{\star}_{W}$) \\
\midrule
$4.3385\times10^{-3}$ & 16  & 6  & 4 \\
$10^{-3}$               & 32  & 10  & 4 \\
$10^{-4}$               & 101 & 21 & 4 \\
$10^{-5}$               & 317 & 46 & 4 \\
\midrule
Asymptotic behavior & $p_0^{-1/2}$ & $p_0^{-1/3}$ & $\Theta(1)$ \\
\bottomrule
\end{tabular}
\end{table}

Table~\ref{tab:finite-prevalence-renewal} translates the scaling results into finite-prevalence numerical decisions. Because only the ratio $\lambda/\mu$ matters for these
optimizers, the same normalized capacity ratio $\lambda/\mu=2$ is used
throughout. Under the classical testing objective, the optimal pool grows
rapidly as prevalence falls, from $16$ at the hospital prevalence to $317$ at
$p_0=10^{-5}$, closely reflecting the square-root law. The queueing-optimal
pool grows more slowly, from $6$ to $46$, in line with
Theorem~\ref{thm:global-queue-cuberoot}.

When all three turnaround components
are minimized under this normalization, the computed numerical minimizer remains $k=4$ at
all four prevalence values. This last column is a finite-parameter observation,
not the statement of Theorem~\ref{thm:bounded-formation}; it illustrates the
same operational force behind that theorem, namely that waiting to assemble
larger pools can offset the statistical benefit of ever lower prevalence.

Taken together, the results in this section provide a decision map for the paper. The square-root rule is the correct benchmark when tests are the scarce resource. The cube-root rule emerges when queueing consequences are priced. A positive formation cost caps routine pool size altogether. The different laws therefore identify different operational environments rather than different approximations of one universal rule.
\begin{remark}
We now briefly discuss the numerical evaluation in Table \ref{tab:finite-prevalence-renewal}. For each prevalence value $p_0$ and each candidate stable integer $k$ (i.e., for fixed $\lambda/\mu=2$, we consider every integer pool size $1\leq k\le K_{\max}=400$ satisfying the stability condition $\rho_k<1$. Here, the testing component, the queueing component, and the total-turnaround time are all evaluated over $1\leq k\leq K_{\max}=400$), we compute the stationary distribution of the embedded queue numerically, by using the finite-state approximation $Q_{n+1}^{(N)}=\min\left\{(Q_n^{(N)}-k)^+ +A_{n+1}^{(k)},N\right\}$, with $N=k+500$. Thus, probability that would lead to a queue length larger than \(N\) is assigned to the boundary state \(N\). Iteration of the corresponding transition equation gives the stationary probabilities $\pi_0^{(N)},\ldots,\pi_N^{(N)}$ of the truncated chain.

These probabilities are then substituted into the finite-state approximations of $\E[W_{\rm idle}(k)]$, and $\E[W_{\rm queue}(k)]$. The in-testing component is $\E[W_{\rm test}(k)]=\frac1\mu+\frac{p_k(k+1)}{2\mu}$, and the total turnaround time is obtained from $\E[W(k)]=\E[W_{\rm idle}(k)]+\E[W_{\rm queue}(k)]+\E[W_{\rm test}(k)]$. The entries of Table \ref{tab:finite-prevalence-renewal} are the numerically computed integer minimizers
over this stable search set $1\le k\le400$. The stationary distribution is computed using the finite-state approximation with $N=k+500$. Repeating the calculations with $N=k+700$ leaves all the derived minimizers unchanged. This comparison is a numerical sensitivity check rather than
a formal truncation-error bound.
\end{remark}

\section{\texorpdfstring{Conclusion}{Conclusion}}
\label{sec:conclusion}

Classical group testing asks how many specimens should be pooled in order to
save tests. Once pooled testing is embedded in a capacity-constrained service
system, that question is incomplete. Pool size also determines how long a
laboratory waits before starting, how much stochastic workload a positive pool
creates, and how strongly that variability is amplified by congestion. The
result is not one universal correction to Dorfman's rule, but a set of
operational regimes tied to different scarce resources.

The classical square-root law remains the correct benchmark when testing
effort is the objective. It also survives, with a congestion-dependent leading
constant, under the unique joint prevalence--capacity scaling that keeps
utilization nondegenerate at the Dorfman scale. When the objective prices the busy-testing component $W_{\rm queue}$ instead, the rare positive-pool workload enters through the
service-time second moment and the relevant scale becomes a cube root. Adding a
fixed positive testing cost changes the cube-root constant but not the
exponent. By contrast, any positive cost of forming batches prevents the
routine fixed-pool recommendation from diverging as prevalence vanishes. These
results explain why operationally appropriate pools can be much smaller than
the classical test-minimizing recommendation, particularly at very low
prevalence.

The state-dependent $(l,k)$ policy gives this classification a practical
interpretation. The lower threshold $l$ determines when the laboratory normally
releases a batch, while the upper threshold $k$ determines how aggressively it
can absorb backlog. In the delay-sensitive low-prevalence regime, the system
selects $B=l$ with probability tending to one and the cube-root law therefore
governs routine release. The congestion cap remains a separate control. Fixed
batching is useful because it reveals the mechanism cleanly; dynamic batching
shows how that mechanism should be implemented.

The numerical exhibits are designed to connect these asymptotic statements to
finite-prevalence decisions. The turnaround decomposition shows which delay
component makes additional pooling unattractive. The prevalence--capacity map
identifies where the congestion-corrected square-root regime is relevant. The
objective comparison shows how quickly the square-root, cube-root, and bounded
rules emerge and how costly it can be to apply Dorfman's test-minimizing pool
size to a laboratory whose actual objective includes time. These cube-root laws are derived under the sequential Dorfman retesting
protocol. Extending them to parallel retesting, imperfect tests, dilution
effects, or other testing protocols is left for future work.

\section*{Acknowledgements} I. Dimitriou gratefully acknowledges the Empirikion Foundation, Athens, Greece (\href{https://www.empirikion.gr/}{www.empirikion.gr}) for the financial support of this work.

\bibliographystyle{plainnat}
\bibliography{pooledtesting2}

@book{bruneel1993discrete,
  author    = {Bruneel, Herwig and Kim, Byung G.},
  title     = {Discrete-Time Models for Communication Systems Including ATM},
  publisher = {Kluwer Academic Publishers},
  address   = {Boston},
  series    = {The Kluwer International Series in Engineering and Computer Science},
  volume    = {205},
  year      = {1993},
  doi       = {10.1007/978-1-4615-3130-2}
}

@article{janssen2005analytic,
  author  = {Janssen, A. J. E. M. and van Leeuwaarden, J. S. H.},
  title   = {Analytic Computation Schemes for the Discrete-Time Bulk Service Queue},
  journal = {Queueing Systems},
  volume  = {50},
  number  = {2--3},
  pages   = {141--163},
  year    = {2005},
  doi     = {10.1007/s11134-005-0402-z}
}

@article{Cinlar1969,
  author  = {{\c{C}}inlar, Erhan},
  title   = {Markov Renewal Theory},
  journal = {Advances in Applied Probability},
  volume  = {1},
  number  = {2},
  pages   = {123--187},
  year    = {1969},
  doi     = {10.2307/1426216}
}

@article{benjaafar2025congestion,
  author  = {Benjaafar, Saif and Jouini, Oualid and Legros, Benjamin and Lu, Bingnan},
  title   = {Pooled Testing in the Presence of Congestion},
  journal = {Operations Research Letters},
  volume  = {58},
  pages   = {107208},
  year    = {2025},
  doi     = {10.1016/j.orl.2024.107208}
}

@article{aprahamian2019risk,
  author  = {Aprahamian, Hrayer and Bish, Douglas R. and Bish, Ebru K.},
  title   = {Optimal Risk-Based Group Testing},
  journal = {Management Science},
  volume  = {65},
  number  = {9},
  pages   = {4365--4384},
  year    = {2019},
  doi     = {10.1287/mnsc.2018.3138}
}

@article{aprahamian2020robust,
  author  = {Aprahamian, Hrayer and Bish, Douglas R. and Bish, Ebru K.},
  title   = {Optimal Group Testing: Structural Properties and Robust Solutions, with Application to Public Health Screening},
  journal = {INFORMS Journal on Computing},
  volume  = {32},
  number  = {4},
  pages   = {895--911},
  year    = {2020},
  doi     = {10.1287/ijoc.2019.0942}
}

@article{barlev2007bulk,
  author  = {Bar-Lev, Shaul K. and Parlar, Mahmut and Perry, David and Stadje, Wolfgang and {van der Duyn Schouten}, Frank A.},
  title   = {Applications of Bulk Queues to Group Testing Models with Incomplete Identification},
  journal = {European Journal of Operational Research},
  volume  = {183},
  number  = {1},
  pages   = {226--237},
  year    = {2007},
  doi     = {10.1016/j.ejor.2006.09.086}
}

@inproceedings{bhimaraju2022scheduling,
  author    = {Bhimaraju, Akhil and Varshney, Lav R.},
  title     = {Scheduling Group Tests over Time},
  booktitle = {2022 IEEE International Symposium on Information Theory (ISIT)},
  pages     = {886--891},
  year      = {2022},
  publisher = {IEEE},
  doi       = {10.1109/ISIT50566.2022.9834434}
}

@article{bish2021robust,
  author  = {Bish, Douglas R. and Bish, Ebru K. and El-Hajj, Hussein and Aprahamian, Hrayer},
  title   = {A Robust Pooled Testing Approach to Expand {COVID-19} Screening Capacity},
  journal = {PLOS ONE},
  volume  = {16},
  number  = {2},
  pages   = {e0246285},
  year    = {2021},
  doi     = {10.1371/journal.pone.0246285}
}

@article{bish2024multidisease,
  author  = {Bish, Douglas R. and Bish, Ebru K. and El Hajj, Hussein},
  title   = {Disease Bundling or Specimen Bundling? Cost- and Capacity-Efficient Strategies for Multidisease Testing with Genetic Assays},
  journal = {Manufacturing \& Service Operations Management},
  volume  = {26},
  number  = {1},
  pages   = {95--116},
  year    = {2024},
  doi     = {10.1287/msom.2022.0296}
}

@article{elhajj2022newborn,
  author  = {El Hajj, Hussein and Bish, Douglas R. and Bish, Ebru K. and Kay, Denise M.},
  title   = {Novel Pooling Strategies for Genetic Testing, with Application to Newborn Screening},
  journal = {Management Science},
  volume  = {68},
  number  = {11},
  pages   = {7994--8014},
  year    = {2022},
  doi     = {10.1287/mnsc.2021.4289}
}

@article{wein1996dilution,
  author  = {Wein, Lawrence M. and Zenios, Stefanos A.},
  title   = {Pooled Testing for {HIV} Screening: Capturing the Dilution Effect},
  journal = {Operations Research},
  volume  = {44},
  number  = {4},
  pages   = {543--569},
  year    = {1996},
  doi     = {10.1287/opre.44.4.543}
}

@article{aldridge2019information,
  author  = {Aldridge, Matthew and Johnson, Oliver and Scarlett, Jonathan},
  title   = {Group Testing: An Information Theory Perspective},
  journal = {Foundations and Trends in Communications and Information Theory},
  volume  = {15},
  number  = {3--4},
  pages   = {196--392},
  year    = {2019},
  doi     = {10.1561/0100000099}
}

@article{abdalhamid2020pooling,
  author  = {Abdalhamid, Baha and Bilder, Christopher R. and McCutchen, Emily L. and Hinrichs, Steven H. and Koepsell, Scott A. and Iwen, Peter C.},
  title   = {Assessment of Specimen Pooling to Conserve {SARS-CoV-2} Testing Resources},
  journal = {American Journal of Clinical Pathology},
  volume  = {153},
  number  = {6},
  pages   = {715--718},
  year    = {2020},
  doi     = {10.1093/ajcp/aqaa064}
}

@book{chaudhry1983bulk,
  author    = {Chaudhry, M. L. and Templeton, J. G. C.},
  title     = {A First Course in Bulk Queues},
  publisher = {Wiley},
  address   = {New York},
  year      = {1983},
  isbn      = {0-471-86260-6}
}

@article{Cabrera2023,
  author = {Cabrera, Ana and Al Mutawah, Fatimah and Kadour, Mike and
            Schofield, Shannon and Conkey, Beverley and Fuller, Jeffrey and
            Payne, Michael and Elsayed, Sameer and Delport, Johan},
  title = {Increasing {SARS-CoV-2} Testing Capacity Through Specimen Pooling: An Acute Care Center Experience},
  journal = {PLOS ONE},
  volume = {18},
  number = {6},
  pages = {e0287137},
  year = {2023},
  doi = {10.1371/journal.pone.0287137}
}

@article{van2003moment,
  title={Moment inequalities for the discrete-time bulk service queue},
  author={Denteneer, D. and Janssen, A.J.E.M. and van Leeuwaarden, J.S.H.},
  journal={Mathematical Methods of Operations Research},
  volume={61},
  number={1},
  pages={85--108},
  year={2005},
  publisher={Springer}
}

@article{dorfman1943,
  author  = {Dorfman, Robert},
  title   = {The Detection of Defective Members of Large Populations},
  journal = {The Annals of Mathematical Statistics},
  year    = {1943},
  volume  = {14},
  number  = {4},
  pages   = {436--440},
  doi     = {10.1214/aoms/1177731363}
}

@article{mutesa2021,
  author  = {Mutesa, Leon and Ndishimye, Patrick and Butera, Yves and Souopgui, Joseph and Uwineza, Aimable and Rutayisire, Richard and Musoni, Eric and Rujeni, Nadine and Nyatanyi, Thierry and Ntagwabira, Edouard and others},
  title   = {A Pooled Testing Strategy for Identifying SARS-CoV-2 at Low Prevalence},
  journal = {Nature},
  year    = {2021},
  volume  = {589},
  number  = {7841},
  pages   = {276--280},
  doi     = {10.1038/s41586-020-2885-5}
}

@article{cleary2021,
  author  = {Cleary, Brian and Hay, James A. and Blumenstiel, Brendan and Harden, Maegan and Cipicchio, Michelle and Bezney, Jon and Simonton, Brooke and Hong, David and Senghore, Madikay and Sesay, Abdul K. and others},
  title   = {Using Viral Load and Epidemic Dynamics to Optimize Pooled Testing in Resource-Constrained Settings},
  journal = {Science Translational Medicine},
  year    = {2021},
  volume  = {13},
  number  = {589},
  pages   = {eabf1568},
  doi     = {10.1126/scitranslmed.abf1568}
}

@article{barak2021,
  author  = {Barak, Netta and Ben-Ami, Roni and Sido, Tal and Perri, Amir and Shtoyer, Aviad and Rivkin, Mila and Licht, Tamar and Peretz, Ayelet and Magenheim, Judith and Fogel, Irit and others},
  title   = {Lessons from Applied Large-Scale Pooling of 133,816 SARS-CoV-2 RT-PCR Tests},
  journal = {Science Translational Medicine},
  year    = {2021},
  volume  = {13},
  number  = {589},
  pages   = {eabf2823},
  doi     = {10.1126/scitranslmed.abf2823}
}

@article{zismanov2024,
  author  = {Zismanov, Shosh and Shalem, Bar and Margolin-Miller, Yulia and Rosin-Grunewald, Dalia and Adar, Roy and Keren-Naus, Ayelet and Amichay, Doron and Ben-Dor, Anat and Shemer-Avni, Yonat and Porgador, Angel and Shental, Noam and Hertz, Tomer},
  title   = {High Capacity Clinical SARS-CoV-2 Molecular Testing Using Combinatorial Pooling},
  journal = {Communications Medicine},
  year    = {2024},
  volume  = {4},
  number  = {1},
  pages   = {121},
  doi     = {10.1038/s43856-024-00531-w}
}

@article{pagotto2025,
  title={Assessing the potential for specimen pooling to streamline nosocomial surveillance of methicillin-resistant Staphylococcus aureus (MRSA)},
  author={Pagotto, Isabella and Alqahtani, Mohammed and Joy, Bryn and McCracken, Gregory R and Davis, Ian R and LeBlanc, Jason J and Patriquin, Glenn},
  journal={Microbiology Spectrum},
  volume={14},
  number={1},
  pages={e02585--25},
  year={2025},
  publisher={American Society for Microbiology 1752 N St., NW, Washington, DC}
}

@article{long2026distributionally,
  title={Distributionally robust group testing with correlation information},
  author={Long, Daniel Zhuoyu and Qi, Jin and Sun, Yu and Zhang, Aiqi},
  journal={Management Science},
  year={2026},
  publisher={INFORMS}
}

@article{augenblick2022pooled,
  title={Pooled testing efficiency increases with test frequency},
  author={Augenblick, Ned and Kolstad, Jonathan and Obermeyer, Ziad and Wang, Ao},
  journal={Proceedings of the National Academy of Sciences},
  volume={119},
  number={2},
  pages={e2105180119},
  year={2022},
  publisher={National Academy of Sciences}
}

@article{lee2022,
  title={Real-world evaluation of a sample pooling strategy for large-scale rapid COVID-19 testing},
  author={Lee, Hao-Ru and Wang, Fang-Yu and Li, Jia-Ling and Chou, Teh-Ying and Ho, Hsiang-Ling},
  journal={Journal of Clinical Virology},
  volume={149},
  pages={105133},
  year={2022},
  publisher={Elsevier}
}

\appendix

\section{Further numerical results}
\label{app:further-numerics}

This appendix provides additional finite-prevalence calculations that complement
the asymptotic results in the main text. The purpose is twofold. First, we
compare the pool sizes selected by three distinct operational criteria: testing
effort, queueing delay, and total turnaround time. Second, we examine how these
choices change with the available testing capacity. This comparison is useful
because the three objectives respond to prevalence and capacity in markedly
different ways.

At low prevalence, the numerical results closely reflect the three regimes
developed in the paper. The testing-effort optimizer follows the classical
square-root scale, whereas the queueing-delay optimizer follows the cube-root
scale. By contrast, the total-turnaround optimizer remains bounded. In the
experiments below, the latter is driven mainly by the capacity ratio and, over
the displayed low-prevalence range, follows the simple numerical pattern
$k_W^\star\approx 2\lambda/\mu$. This should not be interpreted as a new
asymptotic formula; rather, it illustrates the bounded-pooling conclusion of
Corollary~\ref{cor:bounded-turnaround} and the fact that formation time prevents
total-turnaround minimization from favoring increasingly large pools as
prevalence vanishes.

Table~\ref{tab:further-numerics} consolidates all numerical results. The quantities
$k_T^\star$, $k_Q^\star$, and $k_W^\star$ minimize, respectively, the testing
effort $T(k)$, the renewal-reward fixed-batch queueing delay
$\E[W_{\rm queue}]$, and the total turnaround time $\E[W]$. The corresponding
utilizations show how close the selected policies operate to the stability
boundary. For each parameter configuration, we minimize over integer pool sizes
$1\leq k\leq K_{\max}=400$, retaining only stable choices satisfying
$\rho_k<1$. The repeated asymptotic-order rows from the separate tables are
omitted: as established in the main text, $k_T^\star\sim p_0^{-1/2}$,
$k_Q^\star\sim p_0^{-1/3}$, and $k_W^\star=\Theta(1)$.

The final panel considers the more moderate prevalence level $p_0=0.1$. Here
capacity affects not only the preferred pool size but also feasibility. For the
capacity ratios displayed, stable fixed-batch policies exist for
$\lambda/\mu=0.5,1,$ and $1.5$, whereas no stable fixed-batch policy is found
for $\lambda/\mu\ge 2$. Thus, once prevalence is no longer very small,
increasing the pool size need not restore stability: larger pools are positive
more often and consequently generate more frequent and longer retesting
cycles.

\begin{table}[!ht]
\caption{\textbf{Finite-prevalence optimal pool sizes and corresponding utilizations.}
Panel A reports the complete low-prevalence calculations for the five capacity
ratios used in the numerical study. Panel B reports the moderate-prevalence case
$p_0=0.1$, where a dash indicates that no stable fixed-batch policy exists among
the admissible pool sizes. The three optimizer columns illustrate, respectively,
the square-root, cube-root, and bounded-pooling regimes; the utilization columns
show the load induced by each selected policy.}
\centering
\label{tab:further-numerics}
\begin{tabular}{c c c c c c c c}
\toprule
Capacity ratio & Prevalence $p_0$ & $k_T^{\star}$ & $k_Q^{\star}$ & $k_W^{\star}$ & $\rho_{k_T^\star}$ & $\rho_{k_Q^\star}$ & $\rho_{k_W^\star}$ \\
\midrule
\multicolumn{8}{l}{\textit{Panel A: Low-prevalence calculations}}\\
\addlinespace[0.3em]
$\lambda/\mu=0.5$ & $4.3385\times10^{-3}$ & 16 & 6 & 1 & 0.065 & 0.096 & 0.502\\
& $10^{-3}$ & 32 & 10 & 1 & 0.031 & 0.055 & 0.500\\
& $10^{-4}$ & 101 & 21 & 1 & 0.010 & 0.025 & 0.500\\
& $10^{-5}$ & 317 & 46 & 1 & 0.003 & 0.011 & 0.500\\
\addlinespace[0.3em]
$\lambda/\mu=2$ & $4.3385\times10^{-3}$ & 16 & 6 & 4 & 0.259 & 0.385 & 0.534\\
& $10^{-3}$ & 32 & 10 & 4 & 0.126 & 0.220 & 0.508\\
& $10^{-4}$ & 101 & 21 & 4 & 0.040 & 0.099 & 0.501\\
& $10^{-5}$ & 317 & 46 & 4 & 0.013 & 0.044 & 0.500\\
\addlinespace[0.3em]
$\lambda/\mu=3$ & $4.3385\times10^{-3}$ & 16 & 7 & 6 & 0.389 & 0.519 & 0.577\\
& $10^{-3}$ & 32 & 10 & 6 & 0.188 & 0.330 & 0.518\\
& $10^{-4}$ & 101 & 21 & 6 & 0.060 & 0.149 & 0.502\\
& $10^{-5}$ & 317 & 46 & 6 & 0.019 & 0.067 & 0.500\\
\addlinespace[0.3em]
$\lambda/\mu=4$ & $4.3385\times10^{-3}$ & 16 & 8 & 8 & 0.519 & 0.637 & 0.637\\
& $10^{-3}$ & 32 & 11 & 8 & 0.251 & 0.407 & 0.532\\
& $10^{-4}$ & 101 & 21 & 8 & 0.080 & 0.199 & 0.503\\
& $10^{-5}$ & 317 & 45 & 8 & 0.025 & 0.091 & 0.500\\
\addlinespace[0.3em]
$\lambda/\mu=5$ & $4.3385\times10^{-3}$ & 16 & 10 & 10 & 0.649 & 0.713 & 0.747\\
& $10^{-3}$ & 32 & 12 & 10 & 0.314 & 0.476 & 0.550\\
& $10^{-4}$ & 101 & 21 & 10 & 0.100 & 0.249 & 0.505\\
& $10^{-5}$ & 317 & 45 & 10 & 0.032 & 0.113 & 0.500\\
\midrule
\multicolumn{8}{l}{\textit{Panel B: Moderate prevalence, $p_0=0.1$}}\\
\addlinespace[0.3em]
$\lambda/\mu=0.5$ & $0.1$ & 4 & 2 & 1 & 0.297 & 0.345 & 0.550\\
$\lambda/\mu=1$ & $0.1$ & 4 & 3 & 2 & 0.594 & 0.604 & 0.690\\
$\lambda/\mu=1.5$ & $0.1$ & 4 & 3 & 3 & 0.891 & 0.906 & 0.906\\
$\lambda/\mu=2$ & $0.1$ & -- & -- & -- & -- & -- & --\\
$\lambda/\mu=3$ & $0.1$ & -- & -- & -- & -- & -- & --\\
$\lambda/\mu=4$ & $0.1$ & -- & -- & -- & -- & -- & --\\
\bottomrule
\end{tabular}%
\end{table}

Finally, Table \ref{tab:consequences-queueing-optimizers} illustrates that the obtained optimizers in Table \ref{tab:finite-prevalence-renewal} are consistent with the theoretical result established in  Theorem~\ref{thm:global-queue-cuberoot}.
\begin{table}[ht!]
    \centering
    \caption{\textbf{Consequences of Theorem} \ref{thm:global-queue-cuberoot}}
    \label{tab:consequences-queueing-optimizers}
    \begin{tabular}{c c c}
      \toprule
Prevalence $p_0$ & $p_0^{-1/3}$ & $k^{\star}_{Q}$ \\
\midrule
$4.3385\times10^{-3}$ & 6.13  & 6\\
$10^{-3}$               & 10  & 10  \\
$10^{-4}$               & 21.54 & 21  \\
$10^{-5}$               & 46.42 & 46 \\
\midrule
    \end{tabular}
    
\end{table}

\section{Proofs of supporting results}
\label{app:proofs}

This appendix collects the longer proofs that support the asymptotic and optimization results in the main text. The main text retains a proof sketch for each result, emphasizing the mechanism used in the argument, while the subsections below give the complete derivations. The proofs are presented in the order in which the corresponding results appear in the paper. No approximation or intermediate step used in the original arguments is omitted.

\subsection{Proposition~\ref{prop:exact-heavy-traffic}: Exact fixed-batch heavy-traffic limit}
\label{app:proof-exact-heavy-traffic}

\proof By \eqref{eq:fixed-dorfman-inside-root-wait},
\[
\E[W_{\rm queue}(k)]
=
\frac1\lambda\sum_{m=1}^{k-1}\frac1{1-\xi_m}
+
\frac{\lambda \E[(S^{(k)})^2]-(k-1)\E[S^{(k)}]}
     {2(k-\lambda\E[S^{(k)}])}-\E[W_{\rm idle}(k)].
\]
For fixed $k$, the idle contribution $\E[W_{\rm idle}(k)]$ remains bounded as $\rho_k\uparrow1$. Hence, subtracting this term does not affect the leading heavy-traffic singularity. For the roots $\xi_1,\ldots,\xi_{k-1}$ inside the unit disk, the moment bound of \cite[Eq.~(12)]{van2003moment} gives, uniformly for
$\lambda\E[S^{(k)}]<k$,
\[
\sum_{m=1}^{k-1}\frac1{1-\xi_m}
\leq
\frac{k-1}{2}
+
\frac12\min\{\lambda\E[S^{(k)}],k-1\}
\leq k-1.
\]
Hence the root sum remains bounded as $\rho_k\uparrow1$.  Since then
$\lambda\uparrow k/\E[S^{(k)}]>0$, its prefactor $1/\lambda$ also remains
bounded.  Combining this observation with
\eqref{eq:exact-ht-explicit-term} gives
\[
\E[W_{\rm queue}(k)]
=O(1)
+
\frac{\E[S^{(k)}]}{2}
\left[
\frac{\rho_k c_S^2(k)+1/k}{1-\rho_k}-1
\right].
\]
The fraction diverges as $\rho_k\uparrow1$, while $\rho_k\to1$, which yields
\eqref{eq:exact-heavy-traffic}.
\endproof

\subsection{Lemma~\ref{lem:fixedBSQ}: Fixed-batch queueing order}
\label{app:proof-fixed-batch-order}

\proof
From \eqref{eq:fixed-dorfman-wait-boundary}, after subtracting the idle contribution term, the busy-testing contribution is
\begin{equation}
\E[W_{\rm queue}(k)]
=
\frac1k\left[
\E[S^{(k)}]\sum_{i=k}^\infty\pi_i(i-k)
+\frac\lambda2\E[(S^{(k)})^2]
\right].
\label{eq:fixed-boundary-proof-start}
\end{equation}
Because $p_k=1-(1-p_0)^k\sim kp_0$ and $k^2p_0\to0$,
$\E[S^{(k)}]=\frac{1+kp_k}{\mu}\sim\frac1\mu$.
Moreover,
\[
\E[(S^{(k)})^2]
=
\frac{2(1-p_k)+p_k(k+1)(k+2)}{\mu^2}
=
\Theta(1+k^3p_0),
\]
so
\begin{equation}
\frac{\lambda}{2k}\E[(S^{(k)})^2]
=
\Theta\!\left(\frac1k+k^2p_0\right).
\label{eq:fixed-leading-second-moment}
\end{equation}
It remains to control the backlog term.  Let
$L_n=(Q_n-k)^+$.  In the fixed-batch model,
$L_{n+1}=(L_n+A_{n+1}^{(k)}-k)^+$,
which is a Lindley recursion.  Kingman's inequality for this recursion gives
$\E[L] \leq \frac{\Var(A^{(k)})}{2(k-\E[A^{(k)}])}$.
Since the number of arrivals during a testing cycle is mixed Poisson,
\[
\Var(A^{(k)})
=
\lambda\E[S^{(k)}]+\lambda^2\Var(S^{(k)})
=
O(1+k^3p_0).
\]
Also $k-\E[A^{(k)}]=k-\lambda\E[S^{(k)}]\sim k$.  Hence
\[
\sum_{i=k}^\infty\pi_i(i-k)
=
\E[L]
=
O\!\left(\frac{1+k^3p_0}{k}\right),
\]
and therefore
\begin{equation}
\frac{\E[S^{(k)}]}{k}
\sum_{i=k}^\infty\pi_i(i-k)
=
O\!\left(\frac1{k^2}+kp_0\right).
\label{eq:fixed-backlog-lower-order}
\end{equation}
Since $k\to\infty$, we have 
\begin{displaymath}
    \frac{1}{k^2}=o\left(\frac{1}{k}\right),\quad kp_0=o\left(k^2p_0\right).
\end{displaymath}
Consequently,
\begin{displaymath}
    \frac{1}{k^2}+kp_0=o\left(\frac{1}{k}+k^2p_0\right).
\end{displaymath}
Hence, the term in \eqref{eq:fixed-backlog-lower-order} is of lower order than
\eqref{eq:fixed-leading-second-moment}.  Substitution in
\eqref{eq:fixed-boundary-proof-start} proves
\eqref{eq:fixed-queue-order}.
\endproof

\subsection{Theorem \ref{thm:exact-cuberoot-fixed}: Exact fixed-batch cube-root law}
\label{app:proof-exact-cuberoot-fixed}
\proof
Under $k\sim x p_0^{-1/3}$,
$kp_0\to0$ and $k^2p_0\to0$. Hence
\[
\E[S^{(k)}]\sim\frac1\mu,
\qquad
\E[(S^{(k)})^2]
\sim
\frac{2+k^3p_0}{\mu^2}
\sim
\frac{2+x^3}{\mu^2}.
\]
Therefore the second-moment term in
\eqref{eq:fixed-boundary-proof-start} satisfies
\[
\frac{\lambda}{2k}\E[(S^{(k)})^2]
\sim
\frac{\lambda}{2\mu^2}
\left(\frac{2+x^3}{x}\right)p_0^{1/3}.
\]
By \eqref{eq:fixed-backlog-lower-order}, the remaining backlog contribution is
$O(p_0^{2/3})$.  This proves \eqref{eq:exact-cuberoot-fixed}.  Minimizing
$g(x)=2/x+x^2$ gives
$g'(x)=-2/x^2+2x$, so the unique minimizer is $x=1$, with $g(1)=3$.
\endproof

\subsection{Lemma~\ref{lem:statedepBSQ}: State-dependent queueing order}
\label{app:proof-state-dependent-order}

\proof
For every admissible batch size $b\in\{l,\ldots,k\}$, the assumption
$l/k\to\gamma>0$ implies $b=\Theta(k)$ uniformly.  Since $bp_0\to0$,
$\E[(S^{(b)})^2]=\Theta(1+b^3p_0)=\Theta(1+k^3p_0)$.
Using the exact reward formula \eqref{eq:general-wait-reward}, and in particular \eqref{eq:general-queue-reward}, the collection
of second-moment terms in its numerator is therefore
$\Theta(1+k^3p_0)$. Since every selected batch size satisfies $l\leq B\leq k$, and $l/k\to\gamma>0$ we have,
\begin{displaymath}
    \frac{2\gamma}{3}k\leq \E[B]\leq k,
\end{displaymath}
for all sufficiently large $k$. Hence $\E[B]=\Theta(k)$. Therefore, after division by $\E[B]$ (having in mind that in stationarity $\E[B]=\lambda\E[C]$), the second-moment contribution is $\Theta\left(\frac1k+k^{2}p_{0}\right)$.

It remains to bound the excess-backlog term. Put $L_n=(Q_n-k)^+$. Since $(Q_n-f(Q_n))^+=(Q_n-k)^+=L_n$, the
state-dependent recursion gives
$L_{n+1} = \left(L_n+A_{n+1}^{(f(Q_n))}-k\right)^+$. In stationarity, 
\begin{displaymath}
    \sum_{i=k}^{\infty}\pi_i(i-k)=\E[(Q-k)^+]=\E[L].
\end{displaymath}
For Dorfman testing, testing times can be coupled so that $S^{(b)}\leq S^{(k)}$ a.s. for $b\leq k$. Using the same Poisson arrival process during service, this implies $A_{n+1}^{(b)}\leq_{\rm st}A_{n+1}^{(k)}$ for every $b\leq k$. By a standard coupling theorem, we may realize these random variables on a common probability space such that $A_{n+1}^{(b)}\leq A_{n+1}^{(k)}$ a.s simultaneously for every $n$, and every $b\in\{l,\ldots,k\}$. Since the state-dependent policy selects $b=f(Q_n)\in\{l,\ldots,k\}$, this gives $A_{n+1}^{(f(Q_n))}\leq A_{n+1}^{(k)}$ a.s. Let $L_0\leq \widehat{L}_0$ and define the fixed-batch Lindley recursion
\[
\widehat L_{n+1}
=
(\widehat L_n+A_{n+1}^{(k)}-k)^+.
\]
Define now the mapping $\phi(x):=(x-k)^{+}$. Then, $L_{n+1}=\phi(L_n+A_{n+1}^{(f(Q_n))})$, and $\widehat{L}_{n+1}=\phi(\widehat{L}_n+A_{n+1}^{(k)})$. Assuming that $L_n\leq \widehat{L}_{n}$ a.s., since $A_{n+1}^{(f(Q_n))}\leq A_{n+1}^{(k)}$ a.s., we have 
\begin{displaymath}
    L_n+A_{n+1}^{(f(Q_n))}\leq \widehat{L}_n+A_{n+1}^{(k)},\quad\text{a.s.}
\end{displaymath}
Since, $\phi(\cdot)$ is non-decreasing we subsequently have 
\begin{displaymath}
   L_{n+1}=\phi(L_n+A_{n+1}^{(f(Q_n))})\leq \phi(\widehat{L}_n+A_{n+1}^{(k)})=\widehat{L}_{n+1}\quad\text{a.s.}.
\end{displaymath}
Thus, by induction $L_n\leq \widehat{L}_n$ a.s. for every $n$. Starting both recursions from zero and passing to their stationary limits (equivalently, using the usual Loynes construction) yields $\E[L]\leq\E[\widehat L]$. The proof of Lemma~\ref{lem:fixedBSQ} therefore yields
\[
\E[L]
\leq\E[\widehat L]
=
O\!\left(\frac{1+k^3p_0}{k}\right).
\]
Because $\E[S^{(k)}]/\E[B]=O(1/k)$, the excess-backlog contribution is
$O(1/k^2+kp_0)$. Since $\frac1{k^2}+kp_0=o\left(\frac1k+k^2p_0\right)$, the excess-backlog contribution is of lower order than the second-moment
contribution. This proves \eqref{eq:state-queue-order}.
\endproof

\subsection{Proposition~\ref{prop:state-collapse}: Collapse to the lower threshold}
\label{app:proof-collapse-lower-threshold}

\proof
The stochastic-dominance argument in Lemma~\ref{lem:statedepBSQ} gives
\[
\E[(Q-f(Q))^+]
=
O\!\left(\frac{1+k^3p_0}{k}\right)
=
O(p_0^{1/3}).
\]
For any selected batch $b\in\{l,\ldots,k\}$, $bp_0\to 0$, $p_b\sim bp_0$,
$bp_b\sim b^2p_0\leq k^2p_0=O(p_0^{1/3})$. Consequently, $\sup_{l\leq b\leq k}\E[S^{(b)}]=O(1)$.  In stationarity,
$Q\stackrel d=(Q-f(Q))^+ + A^{(f(Q))}$,
and hence
$\E[Q] = O(p_0^{1/3})+\lambda\E[S^{(f(Q))}] =O(1)$.
Since $l\sim y p_0^{-1/3}\to\infty$, Markov's inequality gives
$\Pr(Q\geq l)\leq\frac{\E[Q]}{l}=O(p_0^{1/3})$.
The event $\{B\neq l\}$ can occur only when $Q\geq l$, proving the first
claim.  Moreover,
\[
\E[B]
\leq
l\Pr(Q<l)+k\Pr(Q\geq l)
=
l+(k-l)O(p_0^{1/3})
=
l+O(1).
\]
Since $B\geq l$, we also have $\E[B]\geq l$, and therefore $\E[B]\sim l$.
\endproof
\subsection{Theorem \ref{thm:exact-cuberoot-state}: Exact state-dependent cube-root law}
\label{app:proof-exact-cuberoot-state}
\proof
For $b\leq k$,
$p_b\sim bp_0$ and $b^3p_0$ is uniformly bounded on the cube-root scale.
In particular,
\[
\E[(S^{(l)})^2]
\sim
\frac{2+l^3p_0}{\mu^2}
\sim
\frac{2+y^3}{\mu^2}.
\]
Using \eqref{eq:general-wait-reward}, Proposition~\ref{prop:state-collapse},
and $\E[B]\sim l$, the contribution from states $Q<l$ is
\[
\frac\lambda2
\frac{\E[(S^{(l)})^2]}{\E[B]}\Pr(Q<l)
\sim
\frac{\lambda}{2\mu^2}
\left(\frac{2+y^3}{y}\right)p_0^{1/3}.
\]
For $i\leq k$, the second moments are uniformly $O(1)$, while
$\Pr(Q\geq l)=O(p_0^{1/3})$ and $\E[B]=\Theta(p_0^{-1/3})$.  Hence all
second-moment contributions from states $Q\geq l$ are $O(p_0^{2/3})$.
The excess-backlog term is of the same lower order by the bound used in
Lemma~\ref{lem:statedepBSQ}.  This proves \eqref{eq:exact-cuberoot-state}.

It remains to minimize $h(y)=2/y+y^2$. Since
$h'(y)=-2/y^2+2y$ and $h''(y)=4/y^3+2>0$, the unique minimizer is $y=1$ and
$h(1)=3$. The condition $l\leq k$ requires $y\leq x$. Thus, if $x<1$ is fixed, the minimum over $0<y\leq x$ occurs at $y=x$. When $x$ is chosen jointly with $y$, $y=1$ is feasible by taking $x\geq1$. Finally, $l/k\to\gamma$ implies $y/x\to\gamma$, so $y=1$ gives $x=\gamma^{-1}$ and hence, $k\sim\gamma^{-1}p_0^{-1/3}$.
\endproof

\subsection{Theorem~\ref{thm:extendDorfman}: Exact congestion-corrected Dorfman constant}
\label{app:proof-congestion-corrected-dorfman}

\proof
The exact total turnaround time is given by
\eqref{eq:fixed-dorfman-inside-root-total}:
\[
\E[W(k)]
=
\frac1\lambda\sum_{m=1}^{k-1}\frac1{1-\xi_m}
+
\frac{\lambda\E[(S^{(k)})^2]-(k-1)\E[S^{(k)}]}
     {2(k-\lambda\E[S^{(k)}])}
+
\frac1\mu+p_k\frac{k+1}{2\mu}.
\]
Since $p_k\sim kp_0\sim x\varepsilon$,
\[
\E[S^{(k)}]
=\frac{1+kp_k}{\mu}
\sim
\frac{1+x^2}{\lambda c\varepsilon},
\]
and
\[
\E[(S^{(k)})^2]
=\frac{2(1-p_k)+p_k(k+1)(k+2)}{\mu^2}
\sim
\frac{x^3}{(\lambda c)^2\varepsilon^3}.
\]
Hence
\[
\rho_k
=\frac{\lambda\E[S^{(k)}]}{k}
\longrightarrow
\frac{1+x^2}{cx},
\]
so stability is equivalent asymptotically to $x\in(x_-,x_+)$. For the
second explicit term,
\[
\lambda\E[(S^{(k)})^2]
\sim\frac{x^3}{\lambda c^2\varepsilon^3},
\qquad
(k-1)\E[S^{(k)}]=O(\varepsilon^{-2}),
\]
while
\[
2(k-\lambda\E[S^{(k)}])
\sim
\frac{2(cx-1-x^2)}{c\varepsilon}.
\]
Therefore
\begin{equation}
\frac{\lambda\E[(S^{(k)})^2]-(k-1)\E[S^{(k)}]}
     {2(k-\lambda\E[S^{(k)}])}
\sim
\frac{x^3}{2\lambda c\varepsilon^2(cx-1-x^2)}.
\label{eq:scarce-dominant-exact}
\end{equation}
The in-testing term is only $O(\varepsilon^{-1})$. Moreover,
\cite[Eq.~(12)]{van2003moment} gives
\[
\sum_{m=1}^{k-1}\frac1{1-\xi_m}
\leq
\frac{k-1}{2}
+
\frac12\min\{\lambda\E[S^{(k)}],k-1\}
=O(k)=O(\varepsilon^{-1}),
\]
so the root-sum contribution is also $O(\varepsilon^{-1})$. Thus
\eqref{eq:scarce-dominant-exact} is the unique leading-order contribution and
\eqref{eq:exact-scarce-turnaround} follows.

It remains to minimize
\[
J(x)=\frac{x^3}{cx-x^2-1},
\qquad x\in(x_-,x_+).
\]
Differentiation gives
\[
J'(x)=\frac{x^2(2cx-x^2-3)}{(cx-x^2-1)^2}.
\]
The only stationary point in the feasible interval is
$x^\star=c-\sqrt{c^2-3}$. Furthermore,
\[
J''(x)
=
\frac{2x((c^2-1)x^2-3cx+3)}{(cx-x^2-1)^3}>0
\]
on $(x_-,x_+)$ for $c>2$, so this point is the unique minimizer.
\endproof

\subsection{Lemma~\ref{lem:global-queue-lower}: Global queueing lower bound}
\label{app:proof-global-queue-lower}

\proof
The exact renewal-reward expression used in the proof of Lemma~\ref{lem:fixedBSQ} is
\[
\E[W_{\rm queue}(k)]
=
\frac1k\left[
\E[S^{(k)}]\sum_{i=k}^{\infty}\pi_i(i-k)
+\frac{\lambda}{2}\E[(S^{(k)})^2]
\right].
\]
The excess-backlog term is nonnegative, which immediately gives \eqref{eq:queue-global-lower}.

For Dorfman testing,
\[
\E[(S^{(k)})^2]
=
\frac{2(1-p_k)+p_k(k+1)(k+2)}{\mu^2}.
\]
Suppose first that $\E[W_{\rm queue}(k)]\to0$. We want to prove that $k\to\infty$. Suppose that $k$ does not tend to infinity. Then, there is a fixed integer $K$, such that $k\leq K$ for infinitely many values of $p_0$. Since there are only finite values $\{1,\ldots,K\}$, one fixed value, say $k_0$ should arise infinitely often. Therefore, $p_{k_{0}}=1-(1-p_0)^{k_0}$, and since $p_0\downarrow 0$, and $k_0$ fixed, we have $(1-p_0)^{k_0}\to 1$, so that $p_{k_{0}}\to 0$. As a result, $\E[(S^{(k)})^2]\to\frac{2}{\mu^2}$. Thus, by \eqref{eq:queue-global-lower}
\begin{displaymath}
 \E[W_{\rm queue}(k_0)]\geq \frac{\lambda}{2k_0}\E[(S^{(k_0)})^2]\to
\frac{\lambda}{\mu^2 k_0}>0,    
\end{displaymath}
which contradicts $\E[W_{\rm queue}(k)]\to0$. Thus, $k\to\infty$.

We next show that $kp_0\to0$. If not, there is a subsequence and a constant $\delta>0$ such that $kp_0\ge\delta$. Along that subsequence,
\[
p_k=1-(1-p_0)^k
\ge 1-\exp(-kp_0)
\ge 1-\e^{-\delta}>0.
\]
Consequently, the second term in $\E[(S^{(k)})^2]$ and \eqref{eq:queue-global-lower} imply
\[
\E[W_{\rm queue}(k)]
\ge
\frac{\lambda}{2\mu^2}\,p_k\frac{(k+1)(k+2)}{k},
\]
which diverges because $k\to\infty$, a contradiction. Thus $kp_0\to0$.

Once $kp_0\to0$, we have $p_k\sim kp_0$. In particular, for all sufficiently small $p_0$, $p_k\le1/2$ and $p_k\ge (1/2)kp_0$. Substituting these two bounds into \eqref{eq:queue-global-lower} gives
\[
\E[W_{\rm queue}(k)]
\ge
\frac{\lambda}{2\mu^2k}
\left[1+\frac12 kp_0(k+1)(k+2)\right].
\]
After reducing constants, this has the form \eqref{eq:queue-global-two-sided-lower}. The precise values of $a$ and $b$ are immaterial; what matters is that the two competing terms hold simultaneously along every sequence whose queueing delay vanishes.
\endproof

\subsection{Theorem~\ref{thm:global-dorfman}: Classical Dorfman benchmark}
\label{app:proof-global-dorfman}

\proof
Take $\bar k=\lceil p_0^{-1/2}\rceil$. Since $\bar k p_0\to0$,
\[
T(\bar k)
=
\frac1{\bar k}+p_{\bar k}
\sim
\frac1{\bar k}+\bar k p_0
\sim 2\sqrt{p_0}.
\]
Moreover, $\rho_{\bar k}=(\lambda/\mu)(1/\bar k+p_{\bar k})\to0$, so this candidate is stable for all sufficiently small $p_0$. Hence the optimum satisfies $T(k_T^\star)=O(\sqrt{p_0})\to0$.

Because $T(k)\ge1/k$, this forces $k_T^\star\to\infty$. Because $T(k)\ge p_k$, it also forces $p_{k_T^\star}\to0$, and therefore $k_T^\star p_0\to0$. Thus
\[
T(k_T^\star)
=
\frac1{k_T^\star}+k_T^\star p_0\,[1+o(1)].
\]
Write $x_{p_0}=k_T^\star\sqrt{p_0}$. Dividing the last display by $\sqrt{p_0}$ gives
\[
\frac{T(k_T^\star)}{\sqrt{p_0}}
=
\frac1{x_{p_0}}+x_{p_0}+o(1).
\]
The candidate $\bar k$ shows that the left-hand side is asymptotically at most $2$. Since $x+1/x\ge2$, with equality only at $x=1$, every convergent subsequence of $x_{p_0}$ must converge to one. Hence $x_{p_0}\to1$, proving \eqref{eq:domain-testing}.
\endproof

\subsection{Theorem~\ref{thm:global-queue-cuberoot}: Global queueing-optimal scale}
\label{app:proof-global-queue-cuberoot}

\proof
Consider $\bar k=\lceil p_0^{-1/3}\rceil$. This candidate is stable for all sufficiently small $p_0$, because
$\rho_{\bar k}=(\lambda/\mu)(1/\bar k+p_{\bar k})=O(p_0^{1/3})\to0$.
Theorem~\ref{thm:exact-cuberoot-fixed} then gives
$\E[W_{\rm queue}(\bar k)]=O(p_0^{1/3})$. By optimality,
\[
\E[W_{\rm queue}(k_Q^\star)]
\le
\E[W_{\rm queue}(\bar k)]
=
O(p_0^{1/3})
\longrightarrow0.
\]
We may therefore apply Lemma~\ref{lem:global-queue-lower} to the optimizing sequence. For all sufficiently small $p_0$,
\[
O(p_0^{1/3})
\ge
\E[W_{\rm queue}(k_Q^\star)]
\ge
\frac{a}{k_Q^\star}+b p_0(k_Q^\star)^2.
\]
The first lower-bound term implies $k_Q^\star=\Omega(p_0^{-1/3})$, while the second implies $k_Q^\star=O(p_0^{-1/3})$. Combining the two bounds proves \eqref{eq:domain-queueing}.

It remains to identify the limiting constant. The first assertion implies that there exist constants $0<c<C<\infty$ such that, for small $p_0$,
\begin{displaymath}
c\leq k^{\star}_Q(p_0)p_0^{1/3}\leq C,
\end{displaymath}
thus, the optimizer is somewhere on the cube-root scale, or equivalently, the normalized optimizer $x(p_0):=k_{Q}^\star(p_0)p_0^{1/3}$ is bounded away from zero and infinity. Theorem \ref{thm:exact-cuberoot-fixed} implies that for each fixed $x>0$, if $k(p_0)\sim xp_0^{-1/3}$, then the busy-testing component is asymptotic to $J(x)p_0^{1/3}$, where $J(x):=\frac{\lambda}{2\mu^2}(\frac{2}{x}+x^2)$ is uniquely minimized for $x=1$. However, for the global optimizer, the value $x(p_0)$ is not fixed in advance, but varies with $p_0$, and we only know that $x(p_0)\in[c,C]$. Thus, we need to identify all the possible accumulation points of $x(p_0)$ as $p_0\downarrow 0$. To find this limiting constant and to identify the possible limiting values of $x(p_0)$, let $(p_{0,n})_{n\geq1}$ be a sequence such that  $p_{0,n}\downarrow0$, and let, $x_n:=x(p_{0,n})=k_Q^\star(p_{0,n})p_{0,n}^{1/3}$. Since $x_n\in[c,C]$, by Bolzano–Weierstrass theorem, $x_n$ has a convergent subsequence, say $x_{n_j}$ and there is a constant $\bar x\in[c,C]$ such that $x_{n_j}\to\bar{x}$, i.e.,
\begin{displaymath}
    k_Q^\star(p_{0,n_j})p_{0,n_j}^{1/3}\to\bar x.
\end{displaymath}
With the above result we are ready to apply Theorem \ref{thm:exact-cuberoot-fixed}. Hence, the cube-root asymptotic, gives
\begin{displaymath}
    p_{0,n_j}^{-1/3}\E[W_{\rm queue}(k_Q^\star(p_{0,n_j}))]\to J(\bar{x}).
\end{displaymath}
Since $k_Q^\star(p_{0,n_j})$ is a global optimizer, then for any stable sequence, for all sufficiently large $j$, in the cube-root scale, say $k_{n_j}:=\lceil p_{0,n_j}^{-1/3}\rceil$, we have
\begin{displaymath}
    \E[W_{\rm queue}(k_Q^\star(p_{0,n_j}))] \leq \E[W_{\rm queue}(k_{n_j})].
\end{displaymath}
Moreover, $k_{n_j} p_{0,n_j}^{1/3}\to 1$, and Theorem \ref{thm:exact-cuberoot-fixed} yields
\begin{displaymath}
    p_{0,n_j}^{-1/3}\E[W_{\rm queue}(k_{n_j})]\to J(1).
\end{displaymath}
Therefore, we have $J(\bar{x})\leq J(1)$. However, $J(x)$ is uniquely minimized at $x=1$, so also $J(1)\leq J(\bar x)$. Hence, $\bar{x}=1$. Note that initially we considered an arbitrary sequence $(p_{0,n})$. Therefore, every accumulation point of $x(p_0)$ is equal to $1$ and hence
$ k^{\star}_{Q}(p_0)\sim p_0^{-1/3}$.
\endproof
\subsection{Theorem~\ref{thm:global-testing-queueing}: Testing plus queueing}
\label{app:proof-testing-queueing}

\proof
Again choose $\bar k=\lceil p_0^{-1/3}\rceil$. Since $\bar k p_0=O(p_0^{2/3})$,
\[
T(\bar k)
=
\frac1{\bar k}+p_{\bar k}
=
O(p_0^{1/3}).
\]
Theorem~\ref{thm:exact-cuberoot-fixed} gives
$\E[W_{\rm queue}(\bar k)]=O(p_0^{1/3})$. Hence
$J_{TQ}(\bar k)=O(p_0^{1/3})$, and optimality yields the same bound for $J_{TQ}(k_{TQ}^\star)$.

Both components of $J_{TQ}$ are nonnegative. In particular,
\[
\E[W_{\rm queue}(k_{TQ}^\star)]
\le
\frac{J_{TQ}(k_{TQ}^\star)}{c_Q}
=
O(p_0^{1/3})
\longrightarrow0.
\]
Lemma~\ref{lem:global-queue-lower} therefore applies. Using \eqref{eq:queue-global-two-sided-lower},
\[
O(p_0^{1/3})
\ge
\frac{a}{k_{TQ}^\star}+b p_0(k_{TQ}^\star)^2.
\]
The two terms imply, respectively,
$k_{TQ}^\star=\Omega(p_0^{-1/3})$ and
$k_{TQ}^\star=O(p_0^{-1/3})$, proving \eqref{eq:domain-combined}.
\endproof

\subsection{Proposition \ref{prop:weighted-cuberoot-constant}: Cost-dependent constant within the cube-root family}\label{app:proof-weighted-cuberoot-constant}
\proof
For $k\sim xp_0^{-1/3}$, $p_k\sim kp_0$, so
\[
T(k)
=
\frac1k+p_k
\sim
\frac{1}{x}p_0^{1/3},
\]
because $p_k\sim x p_0^{2/3}$ is of smaller order. Theorem~\ref{thm:exact-cuberoot-fixed} gives
\[
\E[W_{\rm queue}(k)]
\sim
\frac{\lambda}{2\mu^2}
\left(\frac2x+x^2\right)p_0^{1/3}.
\]
Adding the two weighted contributions gives the stated limit. Differentiating its coefficient with respect to $x$ yields
\[
-\frac{c_T+c_Q\lambda/\mu^2}{x^2}
+\frac{c_Q\lambda}{\mu^2}x.
\]
The unique zero is the stated $x_{TQ}^\star$, and the second derivative is positive for $x>0$.
\endproof

\subsection{Theorem \ref{thm:bounded-formation}: Positive formation-benchmark cost implies bounded pooling}\label{app:proof-bounded-formation}
\proof
The fixed benchmark $k_0$ is stable for all sufficiently small $p_0$, and by assumption $J_F(k_0)=O(1)$. Therefore optimality gives
$J_F(k_F^\star)\le J_F(k_0)=O(1)$. Since $R_{p_0}$ is nonnegative,
\[
\frac{c_F}{2\lambda}(k_F^\star-1)
=
c_F\E[W_{\rm form}(k_F^\star)]
\le
J_F(k_F^\star)
=
O(1).
\]
Thus $k_F^\star=O(1)$. The trivial lower bound $k_F^\star\ge1$ gives $k_F^\star=\Theta(1)$.
\endproof

\subsection{Corollary \ref{cor:bounded-turnaround}: Bounded pool size under total-turnaround minimization}\label{app:proof-bounded-turnaround}
\proof
    For each $p_0\downarrow 0$ optimization is performed for stable pool size $k$, i.e., such that $\rho_{k}<1$. Equivalently
    \begin{displaymath}
        \rho_{k}=\frac{\lambda}{\mu}(\frac{1}{k}+1-(1-p_0)^k)\to\frac{\lambda}{k\mu}<1,
    \end{displaymath}
    Thus, any fixed integer $k$ satisfying $k>\lambda/\mu$ is stable for all $p_0\downarrow 0$. Now note that for all $k\in\{k\in\mathbb{N}:k>\lambda/\mu\}$ we have
    \begin{displaymath}
        \begin{array}{rl}
             \E[S^{(k)}]=&\frac{1+kp_{k}}{\mu}\to \frac{1}{\mu},\vspace{2mm}\\
\E[(S^{(k)})^2]=&\frac{2(1-p_{k})+p_{k}(k+1)(k+2)}{\mu^2}\to \frac{2}{\mu^2}.
        \end{array}
    \end{displaymath}
Since $k$ is fixed and strictly positive (i.e., in the stability region) $\E[(Q-k)^+]$ is bounded. Taking into account the boundedness of $\E[S^{(k)}]$ and $\E[(S^{(k)})^2]$, we have that
\begin{displaymath}
    \E[W(k)]=\E[W_{\rm idle}(k)]+\E[W_{\rm queue}(k)]+\E[W_{\rm test}(k)]=O(1).
\end{displaymath}

On the other hand,
\begin{displaymath}
    \E[\bar W(k)]\geq\E[W_{\rm form}(k)]=\frac{k-1}{2\lambda}.
\end{displaymath}
As a result,
\begin{displaymath}
   \E[W(k)]=\E[\bar W(k)]+\E[W_{\rm test}(k)]\geq\frac{k-1}{2\lambda}. 
\end{displaymath}
Thus, for $k=k^{\star}_W(p_0)$,
\begin{displaymath}
    \frac{k^{\star}_W(p_0)-1}{2\lambda}\leq \E[W(k^{\star}_W(p_0))]\leq \E[W(k)]=O(1).
\end{displaymath}
Therefore, $k_W^\star(p_0)=O(1)$. Now since $k_W^\star(p_0)\geq 1$, we conclude that $k_W^\star(p_0)=\Theta(1)$. 
\endproof

\subsection{Proposition \ref{prop:formation-limit}: Limiting pool size with testing and formation costs}
\label{app:proof-formation-limit}
\proof
Theorem~\ref{thm:bounded-formation} implies that the optimizing sequence remains in a finite set of integers. For every fixed $k\in\mathcal K_0$, stability holds for all sufficiently small $p_0$ and $p_k\to0$, so
\[
J_{TF}(k)
\longrightarrow
\frac{c_T}{k}+\frac{c_F}{2\lambda}(k-1).
\]
Because only finitely many candidate values of $k$ can be optimal along the bounded sequence, pointwise convergence on this finite set is uniform. Any subsequential limit of $k_{TF}^\star$ must therefore minimize the limiting discrete objective. If that minimizer is unique, the optimizer must equal it eventually.
\endproof

\section{An outside-root representation for the fixed-batch model}
\label{app:outside-roots}

In this section, we provide an alternative representation of the fixed-batch mean delay. It is included for completeness and for comparison with the root-based analysis in \cite{benjaafar2025congestion}. The renewal-reward formulas in Section \ref{subsec:fixed-batch-dorfman-from-general} remain the primary formulas used in the paper.

For the fixed-batch Dorfman model, note that the pgf of the number of arrivals during one testing cycle is
\[
A(z)
=
(1-p_k)\psi(z)+p_k\psi(z)^{k+1},
\qquad
\psi(z)=\frac{\mu}{\mu+\lambda(1-z)}.
\]
Put
\[
h(z):=\frac{\mu+\lambda(1-z)}{\mu}.
\]
Then \(A(z)=(1-p_k)h(z)^{-1}+p_kh(z)^{-(k+1)}\), and the kernel
equation \(z^k=A(z)\) is equivalent to
\[
D(z):=
z^kh(z)^{k+1}
-(1-p_k)h(z)^k-p_k
=0.
\]

Assume the stable regime
\[
\lambda\E[S^{(k)}]<k.
\]
Then \(z=1\) is a simple root of \(D\). Denote the remaining roots
inside the unit disk by \(\xi_1,\ldots,\xi_{k-1}\), and the roots outside
the unit disk by \(z_1,\ldots,z_{k+1}\). Thus,
\[
D(z)
=
C(z-1)
\prod_{m=1}^{k-1}(z-\xi_m)
\prod_{j=1}^{k+1}(z-z_j),
\]
for a non-zero constant \(C\). The logarithmic derivative of this factorization, evaluated at \(z=1\),
gives
\[
\sum_{m=1}^{k-1}\frac{1}{1-\xi_m}
+
\sum_{j=1}^{k+1}\frac{1}{1-z_j}
=
\frac{D''(1)}{2D'(1)}\Rightarrow \sum_{m=1}^{k-1}\frac{1}{1-\xi_m}
=
\sum_{j=1}^{k+1}\frac{1}{z_j-1}
+
\frac{D''(1)}{2D'(1)}.
\]

In the present model, $D'(1)=k-\lambda\E[S^{(k)}]$, and thus,
\[
D''(1)
=
k(k-1)-2k(k+1)\frac{\lambda}{\mu}
+k\bigl(2+(k-1)p_k\bigr)(\frac{\lambda}{\mu})^2.
\]
Using
\[
\E[S^{(k)}]=\frac{1+kp_k}{\mu},
\qquad
\E[(S^{(k)})^2]
=
\frac{2(1-p_k)+p_k(k+1)(k+2)}{\mu^2},
\]
simple algebraic calculations yield
\[
\frac{\lambda\E[(S^{(k)})^2]
-(k-1)\E[S^{(k)}]}
{2(k-\lambda\E[S^{(k)}])}
+
\frac{D''(1)}{2\lambda D'(1)}
=
\frac{k-1}{2\lambda}-\frac{k+1}{\mu}.
\]

Substitution into the inside-root representation
\eqref{eq:fixed-dorfman-inside-root-wait} therefore gives
\[
\E[\bar W]
=
\frac{k-1}{2\lambda}
+
\frac{1}{\lambda}
\sum_{j=1}^{k+1}\frac{1}{z_j-1}
-\frac{k+1}{\mu}.
\]

Finally, adding the testing delay
\[
\tau_k
=
\frac1\mu+p_k\frac{k+1}{2\mu},
\]
we obtain
\[
\E[W]
=
\frac{k-1}{2\lambda}
+
\frac{1}{\lambda}
\sum_{j=1}^{k+1}\frac{1}{z_j-1}
-\frac{k}{\mu}
+p_k\frac{k+1}{2\mu}.
\]

The term \((k-1)/(2\lambda)\) is equal to the batch-formation
benchmark delay \(\E[W_{\mathrm{form}}]\). Clearly, it does not coincide with the
renewal-reward idle component \(\E[W_{\mathrm{idle}}]\). The formulas
above represent the same total fixed-batch delay as the renewal-reward
formulas, but they do not provide the decomposition into idle-formation
and busy-testing rewards.

\begin{lemma}[Phase-type testing times yield a rational kernel]
\label{lem:phase-type-rational}
Consider the fixed-batch model with batch size \(k\). Let \(S_k\) denote the
testing time of a batch of size \(k\), and suppose that \(S_k\) has a
phase-type distribution. Define $A^{(k)}(z)
=
\mathbb E\!\left[e^{-\lambda(1-z)S_k}\right]$. Then, \(A^{(k)}(z)\) is a rational function of \(z\). Consequently, the kernel
$K_k(z) = z^k-A^{(k)}(z)$
is rational as well. More precisely, there exist coprime polynomials
\(P_k\) and \(Q_k\) such that
$A^{(k)}(z) = \frac{P_k(z)}{Q_k(z)}$,
and hence
$K_k(z) = \frac{z^kQ_k(z)-P_k(z)}{Q_k(z)}$.
\end{lemma}

\proof
Since arrivals during a testing period form a Poisson process with rate
\(\lambda\),
\[
A^{(k)}(z)
=
\mathbb E\!\left[e^{-\lambda(1-z)S_k}\right],
\]
which is the Laplace transform of \(S_k\) evaluated at
\(\lambda(1-z)\). The Laplace transform of a phase-type distribution is
rational, and therefore \(A^{(k)}(z)\) is rational. The stated representation
for \(K_k(z)\) follows immediately.
\endproof
\begin{remark}[Rational kernels]
\label{rem:rational-kernel}
Suppose that the assumptions of Lemma~\ref{lem:phase-type-rational} hold.
Rationality implies that the zeros of the kernel are the zeros of the polynomial $D_k(z)=z^kQ_k(z)-P_k(z)$, apart from cancellations already removed in the coprime representation. A finite root representation therefore remains available when this polynomial has the same root separation as in the Dorfman case: a simple zero at $z=1$, $k-1$ zeros inside the unit disk, and no other zeros on the unit circle. Under this condition, the stationary pgf can be factorized in terms of the outside roots and the corresponding mean-waiting representation follows analogously. Thus Lemma~\ref{lem:phase-type-rational} provides the algebraic basis for extending the root method, while the required root location must be verified for the testing-cycle distribution at hand.
\end{remark}

\end{document}